\documentclass[a4paper,10pt]{article}

\usepackage{jheppub} 

\usepackage{graphicx,verbatim,epsfig}
\usepackage{epstopdf}
\usepackage[utf8x]{inputenc}
  \usepackage{bm}
   \usepackage{amsmath}
    \usepackage{amssymb}
     \usepackage{pifont}

\newcommand{\Ds}{\displaystyle}

\newcommand{\nn}{\nonumber}

\newcommand{\Tr}{\mathrm{Tr}}

\newcommand{\Li}{\text{Li}}

\renewcommand{\(}{\left(}
\renewcommand{\)}{\right)}

\renewcommand{\vec}[1]{\bm{#1}}

\title{Soft factors for double parton scattering at NNLO}

\author{Alexey Vladimirov}

\affiliation{Institut f\"ur Theoretische Physik, Universit\"at Regensburg,\\
D-93040 Regensburg, Germany}

\emailAdd{aleksey.vladimirov@gmail.com}

\abstract{
We show at NNLO that the soft factors for double parton scattering (DPS) for both integrated and unintegrated kinematics, can be presented entirely in the terms of the soft factor for single Drell-Yan process, i.e. the transverse momentum dependent (TMD) soft factor. Using the linearity of the logarithm of TMD soft factor in rapidity divergences, we decompose the DPS soft factor matrices into a product of matrices with rapidity divergences in given sectors, and thus, define individual double parton distributions at NNLO. The rapidity anomalous dimension matrices for double parton distributions are presented in the terms of TMD rapidity anomalous dimension. The analysis is done using the generating function approach to web diagrams. Significant part of the result is obtained from the symmetry properties of web diagrams without referring to explicit expressions or a particular rapidity regularization scheme. Additionally, we present NNLO expression for the web diagram generating function for Wilson lines with two light-like directions.
}

\begin{document}
\maketitle
\flushbottom

\section{Introduction}
\label{sec:intro}

The effects of double parton scattering (DPS), i.e. the scattering with two partons of a hadron participating in the hard subprocess, are usually expected to be small in comparison to a single parton scattering contribution. However, at very high energies the effect of multiple parton interactions increases and presents an important part of the total cross section, see e.g. \cite{Abe:1997xk,Alitti:1991rd,Abazov:2009gc}. It is known that DPS processes can form strong background for Higgs searches \cite{DelFabbro:1999tf,Bandurin:2010gn}, as well as, be dominant channel for particular reactions, e.g. in the double Drell-Yan process $p+p\to W^++W^+$ \cite{Gaunt:2010pi}. Therefore, the practical interest to DPS processes is constantly increasing.

From the theoretical side, the DPS processes are studied rather weakly. One of the reasons is the cumbersome kinematic structure of DPS. The double parton distributions (DPDs), the analogs of parton distribution functions for DPS, are functions of many variables: two momentum fractions and three transverse coordinates (or one transverse coordinate in the integrated case), say nothing of dependencies on two factorization scales. Additionally, DPDs have reach polarization and color structure, and even the leading order factorization formula for unpolarized and integrated double Drell-Yan involves more than dozen presumably independent DPDs \cite{Diehl:2011yj,Manohar:2012jr}. Nonetheless, during recent years there was significant progress in the theoretical understanding of DPS processes, due to the formulation of appropriate factorization theorems \cite{Diehl:2011yj,Diehl:2015bca,Manohar:2012jr}. 

Apart from increased number of various functions, the DPS factorization theorems resemble the factorization formula for transverse momentum dependent (TMD) processes, see e.g. \cite{Collins:2011zzd}. It is not accidental since the dominant field modes are the same for TMD processes and DPS processes. This analogy grants the possibility to re-use the TMD experience during consideration of DPDs. For example, at NLO all the evolution properties for DPDs can be presented via corresponding evolution properties of TMD distributions \cite{Snigirev:2003cq,Diehl:2011yj,Manohar:2012jr}. 

In this work, we concentrate on the study of DPS soft factors, which are essential part of DPS factorization theorems. Soft factors represent the underlying  interaction of soft gluons  and contain the mixture of rapidity divergences related to both hadrons. This substructure should be decomposed into the parts with rapidity divergences belonging to a given hadron. Only after such decomposition a finite, i.e. meaningful, parton distributions can be defined. Naturally, the decomposition introduces the rapidity parameter. The anomalous dimension for the rapidity parameter scaling also can be deduced from the soft factor. Therefore, the study of the soft factor is an important part of the study of DPS factorization theorems. 

At NLO the soft factors are nearly trivial objects. This order is given by single gluon exchange diagrams only. Therefore, DPS soft factors at NLO scatter into NLO TMD soft factors \cite{Diehl:2011yj,Manohar:2012jr}. At NNLO many non-trivial aspects of  perturbative expansion  arise. The most important one is that simultaneous interaction of several Wilson lines becomes possible, and thus, one can expect highly interesting dynamics. However, the difficulty of consideration also grows. For example, the properties of TMD soft factor, although known for a long time, have been explicitly demonstrated at NNLO only recently \cite{Echevarria:2015byo}.

The soft factors for DPS are rather involved objects composed of four half-infinite light-like cusps of Wilson lines positioned at four different points in the transverse plane  and connected in all possible ways. Consideration of such an object within a usual diagrammatic is a serious calculation problem, mostly due to confusing combinatoric of the color flow. The structure of perturbative series is exceptionally simplified within the generating function approach for web diagrams, formulated in \cite{Vladimirov:2015fea,Vladimirov:2014wga}. Within this approach, one should calculate the generating function, which is unique for a given geometry (in the case of TMD-like soft factors, the only important point is two light-like directions). Various matrix elements such as TMD soft factor, DPS soft factors, are obtained by a projection operation on the generating function. In this way, the usually difficult diagrammatic combinatoric is reduced to a couple of lines of simple algebraic manipulations.

One of the most attractive features of the generating function approach is an efficient organization of the expression. In particularly, it allows to avoid the calculation of whole sectors of diagrams, showing their equivalence with lower perturbative orders. As we demonstrate in this work, the consideration of the generating function at NNLO immediately shows the possibility to present any DPS soft factor in terms of TMD soft factors at this order. This fact is not trivial, since the NNLO expression contains products of TMD soft factor, but does not contain new functions. On the level of diagrams it implies that diagrams in particular combinations cancel each other, while in other combinations scatter into one-loop integrals. To find such combinations can be a tricky task, but they reveal automatically within the generating function approach.

In this paper we demonstrate that at NNLO DPS soft factors are given by a simple combination of TMD soft factors. Having at hands DPS soft factors at NNLO we study the structure of rapidity divergences and present the rapidity evolution equations at NNLO. It appears that some important results can be obtained without referring to expressions for diagrams. For example, the factorization of rapidity divergences for DPD soft factor at NNLO appears to be direct consequence of the rapidity factorization for TMD soft factor. We also perform the explicit calculation within the $\delta$-regularization scheme \cite{Echevarria:2015byo} and confirm the results of the general analysis. The expression for the NNLO generating function for web diagrams presented here for the first time can be also used in other applications.

In the section \ref{sec:factorization} we review the derivation of factorization formula for double Drell-Yan process following articles \cite{Diehl:2011tt,Diehl:2011yj,Diehl:2015bca,Manohar:2012pe,Manohar:2012jr}. This section is mostly needed to introduce the compact notation and necessary details about DPS soft factors. In the section \ref{sec:generating_function} we give a short introduction to generating function approach for web diagrams. In sections \ref{sec:1loop} and \ref{sec:2loop} we discuss the details of evaluation of the generating function for DPS soft factors at NLO and NNLO, respectively. The particular form of projecting operators for DPS soft factors is given in section \ref{sec:action_of_proj}. In sections \ref{sec:TMD_SF},\ref{sec:S4},\ref{sec:S1} we perform the projection operations and obtain the expression for DPS soft factors in terms of TMD soft factors. The origin of the simple structure of DPS soft factors is discussed in \ref{sec:S4}. In section \ref{sec:DPD_soft} we discuss the influence of NNLO expression on the DPS factorization theorem. In particular, we show the factorization of rapidity divergences and define individual DPDs in sec.\ref{sec:recombine}. The rapidity evolution equations at NNLO  are given in sec.\ref{sec:rap_evol}.  The consideration is done in the most general case of unintegrated DPS. The important case of integrated DPS is obtained from these results and presented in the sec.\ref{sec:integrated}.

Technical details of the evaluation are collected in the set of appendices. In the appendix \ref{app:normalizations} we compare our notation with the notation used in \cite{Diehl:2011yj} and \cite{Manohar:2012jr}. The explicit expressions for diagrams, as well as, their analysis are given in appendix \ref{app:calc}. The expressions for basic loop integrals that participate in the generating function are collected in appendix \ref{app:loop_integrals}.

\section{Factorization of double parton scattering}
\label{sec:factorization}

In this section we review some aspects of the double parton scattering factorization. The main aim of this section is to introduce notation and make connections with previous works. The consideration presented here is very superficial, and is an extraction from \cite{Diehl:2011tt,Diehl:2011yj,Diehl:2015bca,Manohar:2012pe,Manohar:2012jr}, to which we refer for the proofs. To get access to the leading order factorized cross-section we use the SCET II technique. The consideration is in many aspects similar to the consideration of Drell-Yan process at moderate transverse momentum \cite{Manohar:2012jr,Becher:2014oda,Becher:2010tm,GarciaEchevarria:2011rb,Collins:2011zzd} (the so-called TMD factorization). Within this context, our main attention is devoted to the geometry of the double-parton scattering and to the color flow. Thus, we skip many important questions of DPS redirecting the reader to the literature \cite{Diehl:2011tt,Diehl:2011yj,Diehl:2015bca,Manohar:2012pe,Manohar:2012jr}.

\subsection{Leading order factorization for double-Drell-Yan}

The cross-section of double Drell-Yan process is given by the following matrix element \cite{Diehl:2011yj,Manohar:2012jr}
\begin{eqnarray}\label{def:double_DY_cross}
\frac{d\sigma}{d\vec X}&=&d\hat \sigma_{\{\mu\}}\int d^4z_{1,2,3} e^{iq_1\cdot(z_1-z_4)}e^{iq_2\cdot(z_2-z_3)}
\\\nn &&\qquad \langle P_1P_2|\bar T \{J^{\dagger\mu_4}(z_4)J^{\dagger\mu_3}(z_3)\}T\{J^{\mu_2}(z_2)J^{\mu_1}(z_1)\}|P_1P_2\rangle,
\end{eqnarray}
where $d\hat \sigma_{\{\mu\}}$ is a leptonic tensor and $J^\mu\sim\bar q \gamma^\mu q$ is the quark-to-vector boson current. Throughout the text an index enclosed in curly brackets denotes the set of  indices of the same kind, e.g. here $d\hat \sigma_{\{\mu\}}=d\hat\sigma_{\mu_1\mu_2\mu_3\mu_4}$. As usual, we define two light-like vectors $n$ and $\bar n$ along the largest components of $P_1$ and $P_2$ correspondingly, with $n\cdot\bar n=1$. The vector decomposition reads $v^\mu=n^\mu v^-+\bar n^\mu v^++\vec v^\mu$, where $\vec v$ are transverse components ($\vec P_1=\vec P_2=0$) and $\vec v^2>0$. The phase-space element $d\vec X$ denotes the complete phase-space of produced bosons, i.e. 
$$d\vec X=dx_1d\bar x_1d \vec q_1~dx_2d\bar x_2d \vec q_2=s^{-2}dq^2_1dY_1d \vec q_1~dq^2_2dY_2d \vec q_2,$$
with $x_i=q^+_i/P_1^+$, $\bar x_i=q_i^-/P_2^-$ (in the reference frame), $s=(P_1+P_2)^2$ is center-mass-energy and $Y_i=\ln(q_i^+/q_i^-)/2$  is the rapidity of produced boson. The large components of momenta are $q_i^+\sim q_i^-\sim P^+_1\sim P^-_2 \sim Q$, where $Q$ is a generic large scale. One of the coordinates, say $z_4$, can be set to zero due to translation invariance, but we keep it explicit for homogeneity of notation and for later convenience. Also in the following we often use the shorthand notation for set of arguments $(\vec b_1,\vec b_2,\vec b_3,\vec b_4)$ as $\vec b_{1,2,3,4}$ (the order indices is important). In the following, although we introduce the notation convenient for our study, we try to be close to the notation and normalizations of \cite{Diehl:2011tt}.

Integrated double Drell-Yan process attracts even more practical interest. The integrated cross-section has the phase-space element $d X=dx_1d\bar x_1~dx_2d\bar x_2$. It can be obtained from the unintegrated cross-section (\ref{def:double_DY_cross}) by the integration over the transverse momenta $\vec q_{1,2}$. Consequently the expressions for the integrated anomalous dimensions, soft factors and over elements can be obtained from the unintegrated ones. In the following sections we consider only the general case of unintegrated kinematics. The expressions for the integrated case are collected in the section \ref{sec:integrated}.

Following the SCET II factorization procedure we consider the quark field in the background gluon field, separating soft and collinear modes \cite{Bauer:2000yr,Bauer:2001yt,Lee:2006nr,Becher:2014oda},
\begin{eqnarray}\label{quark-field-decomp}
q^a_{j}(z)&=& W^{aa'}_n[z,-\infty] \xi^{a'}_{n,j}(z)+W^{aa'}_{\bar n}[z,-\infty] \xi^{a'}_{\bar n,j}(z),
\\ \nn 
\bar q^a_{j}(z)&=&\bar \xi^{a'}_{n,j}(z)W^{a'a}_n[-\infty,z]+\bar \xi^{a'}_{\bar n,j}(z)W^{a'a}_{\bar n}[-\infty,z],
\end{eqnarray}
where $a,a'$ are color indices, $j$ is spinor index, $W^{aa'}_n[z_1,z_2]$ is a (soft) Wilson line from the point $z_1$ to $z_2$, and field $\xi_n$ is the "large"  component of quark field along vector $n$. The explicit definition of Wilson line is
\begin{eqnarray}\label{def:Wilson-line}
W_n[z_1,z_2]=P\exp\(-ig \int_{z_1}^{z_2} dx n^\mu A_\mu^a(x)t^a\).
\end{eqnarray}

The relation inverse to (\ref{quark-field-decomp}) is obtained by applying corresponding "large-component" projector
\begin{eqnarray}
\xi^a_{n,j}(z)=\widetilde W^{aa'}_n[z,-\infty]P^n_{jj'}q_{j'}^{a'}(z),
\end{eqnarray}
and similar for anti-quark and $\bar n$ components. Here $P^n=\gamma^+ \gamma^-$ is the projector in $n$-direction. The Wilson line $\widetilde W$ has the same formal definition as $W$, but instead of soft gluons it consists of collinear ones. 

Substituting the field decomposition (\ref{quark-field-decomp}) into the matrix element (\ref{def:double_DY_cross}) one obtains a large set of terms. The central point of the SCET approach is that at the leading order of factorization and in the absence of Glauber interaction (which has been proved in \cite{Diehl:2015bca}), the field $\xi$ does not interact with soft-gluons, and soft-gluon can be split up into separate matrix element. Then the cross-section is presented in the form
\begin{eqnarray}\label{SCET_factorized}
\frac{d\sigma}{d\vec X}&=&d\hat \sigma_{\{ij\}}\int d^4z_{1,2,3} e^{iq_1\cdot(z_1-z_4)}e^{iq_2\cdot(z_2-z_3)}\sum_{\bar v_i,v_i=n,\bar n}
\\\nn&&\times \langle P_1P_2|\bar T \{\bar \xi^{a_4}_{\bar v_4,j_4}\xi^{b_4}_{v_4,i_4}(z_4)\bar \xi^{a_3}_{\bar v_3,j_3}\xi^{b_3}_{v_3,i_3}(z_3)\}
T\{\bar \xi^{a_2}_{\bar v_2,j_2}\xi^{b_2}_{v_2,i_2}(z_2)\bar \xi^{a_1}_{\bar v_1,j_1}\xi^{b_1}_{v_1,i_1}(z_1)\}|P_1P_2\rangle
\\\nn&&\times \langle 0|
\bar T\{\Lambda_{\bar v_4 v_4}^{a_4b_4}(z_4)\Lambda_{\bar v_3 v_3}^{a_3b_3}(z_3)\}T\{\Lambda_{\bar v_2 v_2}^{a_2b_2}(z_2)\Lambda_{\bar v_1 v_1}^{a_1b_1}(z_1)\}
|0\rangle+...,
\end{eqnarray}
where we extract the Lorentz structures from currents and absorb them into the tensor $\sigma_{\{ij\}}$. The symbol $\Lambda$ denotes a light-like cusp of half-infinite Wilson lines located at position $z$,
\begin{eqnarray}
\Lambda_{v_1v_2}^{ab}(z)=W^{ac}_{v_1}[-\infty,z]W^{cb}_{v_2}[z,-\infty].
\end{eqnarray}
Note, that $\Lambda_{vv}^{ab}(z)=\delta^{ab}$. The dots in (\ref{SCET_factorized}) denote the terms suppressed by powers of $s$ \cite{Bauer:2001yt,Lee:2006nr}. The hard matching coefficients of the vector currents to SCET fields are hidden inside the function $\sigma_{\{ij\}}$. The separation of the hard part introduces the renomalization scales $\mu$'s for each hard sub-process. In the following hard renormalization scales are taken equal to $\mu$ for brevity.

The further consideration is based on the following assumptions that are correct at the leading order of factorization in the region $\Lambda^2_{QCD}\ll \vec q^2_1\sim \vec q^2_2 \ll s$  \cite{Diehl:2011yj,Manohar:2012jr,Bauer:2001yt,Becher:2014oda,Collins:2011zzd}: 
\begin{itemize}
\item Soft radiation does not resolve collinear scales, therefore, soft Wilson lines can be expanded at light-cone origin, $\Lambda(z)\to \Lambda(\vec b)$, where $\vec b$ is the transverse component of $z$; 
\item The "large" components of quark fields couple only to the hadron with corresponding momentum, i.e. $\xi_n$ couples to hadron $P_1$, while $\xi_{\bar n}$ couples to hadron $P_2$;
\item The "large" component of the quark field does not resolve the scales in perpendicular direction, therefore, it can be expanded in that direction, i.e. $\xi_n(z)\to \xi_n(0,z^-,\vec b)$.
\end{itemize}
Using these assumptions one can compute the leading contribution to double-Drell-Yan process. The factorized expression contains various terms with usual parton distributions, double-parton-distribution and combination that mix with each other within the operator-product expansion (see \cite{Gaunt:2012dd} for the leading order analysis). The separation of these terms from each other is an involved procedure (for theoretical development see \cite{Gaunt:2012dd,Diehl:2016khr,Diehl:2011yj,Manohar:2012pe}). In this article we are interested in the study of soft factors responsible only for multi-parton scattering. Therefore, we skip the discussion on the mixture between various matrix elements and consider only the DPS contribution. 

\begin{figure}[t]
\centering
\includegraphics[width=0.95\textwidth]{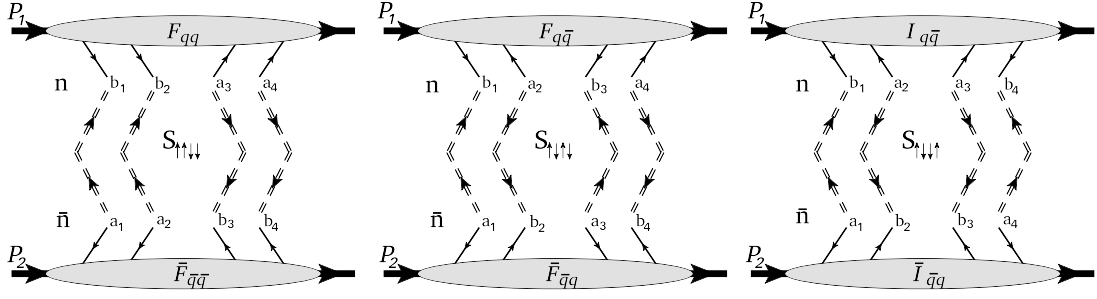}
\caption{Graphical representation of contributions into the DPS factorization formula (\ref{DPS_cross_section}). The first terms in lines of (\ref{DPS_cross_section}) are presented. The out-going (incoming) to blobs fermion line represents quark (antiquark) field $\xi$ ($\bar \xi$). The double-lines represents the soft Wilson lines.} \label{fig:crosssection_terms}
\end{figure}

The DPS part of the cross-section corresponds to terms with simultaneous radiation of two distinct partons from the hadron. Applying leading order factorization restrictions to expression (\ref{SCET_factorized}) and extracting the DPS contributions we obtain
\begin{eqnarray}\label{DPS_cross_section}
\frac{d\sigma}{d\vec X}\Big|_{\text{DPS}}&=&d\hat \sigma_{\{ij\}}\int d^2\vec b_{1,2,3} e^{-i\vec{q}_1\cdot(\vec{b}_1-\vec{b}_4)}e^{-i\vec{q}_2\cdot(\vec{b}_2-\vec{b}_3)}\Bigg[
\\\nn && F^{b_1b_2a_3a_4}_{qq,\{ij\}}\bar F^{a_1a_2b_3b_4}_{\bar q\bar q,\{ij\}}S^{\{ab\}}_{\uparrow\uparrow\downarrow\downarrow}+
F^{a_1a_2b_3b_4}_{\bar q\bar q,\{ij\}}\bar F^{b_1b_2a_3a_4}_{qq,\{ij\}}S^{\{ab\}}_{\downarrow\downarrow\uparrow\uparrow}
\\\nn&&
F^{b_1a_2b_3a_4}_{q\bar q,\{ij\}}\bar F^{a_1b_2a_3b_4}_{\bar q q,\{ij\}}S^{\{ab\}}_{\uparrow\downarrow\uparrow\downarrow}+
F^{a_1b_2a_3b_4}_{\bar q q,\{ij\}}\bar F^{b_1a_2b_3a_4}_{q\bar q,\{ij\}}S^{\{ab\}}_{\downarrow\uparrow\downarrow\uparrow}
\\\nn&&
\mathcal{I}^{b_1a_2a_3b_4}_{ q\bar q,\{ij\}}\bar{\mathcal{I}}^{a_1b_2b_3a_4}_{\bar q q,\{ij\}}S^{\{ab\}}_{\uparrow\downarrow\downarrow\uparrow}+
\mathcal{I}^{a_1b_2b_3a_4}_{\bar qq,\{ij\}}\bar{\mathcal{I}}^{b_1a_2a_3b_4}_{q\bar q,\{ij\}}S^{\{ab\}}_{\downarrow\uparrow\uparrow\downarrow}
\Bigg],
\end{eqnarray}
here we have suppressed arguments of functions for brevity. The functions $S$ are soft factors and given by expressions
\begin{eqnarray}\label{SF_gen_def}
S^{\{ab\}}_{\bullet_1\bullet_2\bullet_3\bullet_4}(\vec b_1,\vec b_2,\vec b_3,\vec b_4)=
\langle 0|
\bar T\{\Lambda_{\bullet_4}^{a_4b_4}(\vec b_4)\Lambda_{\bullet_3}^{a_3b_3}(\vec b_3)\}T\{\Lambda_{\bullet_2}^{a_2b_2}(\vec b_2)\Lambda_{\bullet_1}^{a_1b_1}(\vec b_1)\}
|0\rangle,
\end{eqnarray}
where $\bullet=\downarrow=\bar n n,$ and $\bullet=\uparrow=n \bar n$ (note, that order of arguments and indices of the function $S$ is opposite to their order in the matrix element, i.e. graphical). The ``arrow'' notation is the visual representation of color-flow between light-cone infinities, i.e. if one writes all indices related to $n \infty$ as down indices and all indices related $\bar n \infty$ as up indices, the arrows indicate the order of connection, see fig.\ref{fig:crosssection_terms}. The functions $F$ are double parton distributions (DPDs) and given by expressions
\begin{eqnarray}\label{DPD_TMD_def1}
F_{qq,\{ij\}}^{b_1b_2a_3a_4}(x_{1,2},\vec b_{1,2,3,4})&\simeq& \int dz_{1,2,3}^-e^{x_1 P_1^+(z_1^--z_4^-)}e^{ix_2 P_1^+(z_2^--z_3^-)}
\\&&\nn 
\times \langle P_1|
\bar T\{\bar \xi^{a_4}_{n,j_4}(z_4)\bar \xi^{a_3}_{n,j_3}(z_3)\} T\{\xi^{b_2}_{n,i_2}(z_2)\xi^{b_1}_{n,i_1}(z_1)\}
|P_1\rangle\Big|_{z^+_i=0},
\\\label{DPD_TMD_def2}
F_{q\bar q,\{ij\}}^{b_1a_2b_3a_4}(x_{1,2},\vec b_{1,2,3,4})&\simeq& \int dz_{1,2,3}^-e^{ix_1 P_1^+(z_1^--z_3^-)}e^{ix_2 P_1^+(z_2^--z_4^-)}
\\&&\nn 
\times \langle P_1|
\bar T\{\bar \xi^{a_4}_{n,j_4}(z_4)\xi^{b_3}_{n,i_3}(z_3)\} T\{\bar \xi^{a_2}_{n,i_2}(z_2)\xi^{b_1}_{n,j_1}(z_1)\}
|P_1\rangle\Big|_{z^+_i=0},
\\\label{DPD_TMD_def3}
\mathcal{I}_{q\bar q,\{ij\}}^{b_1a_2a_3b_4}(x_{1,2},\vec b_{1,2,3,4})&\simeq& \int dz_{1,2,3}^-e^{ix_1 P_1^+(z_1^--z_4^-)}e^{ix_2 P_1^+(z_2^--z_3^-)}
\\&&\nn 
\times \langle P_1|
\bar T\{\xi^{b_4}_{n,j_4}(z_4)\bar \xi^{a_3}_{n,i_3}(z_3)\} T\{\bar \xi^{a_2}_{n,j_2}(z_2)\xi^{b_1}_{n,i_1}(z_1)\}
|P_1\rangle\Big|_{z^+_i=0}.
\end{eqnarray}
The similarity sign implies possible normalization factor, which depends on the spinor structure. The distributions $\bar F$ are obtained by changing components $\pm\to \mp$, $n\to \bar n$, and hadron states $P_1\to P_2$, e.g
\begin{eqnarray}
\bar F_{qq,\{ij\}}^{a_1a_2b_3b_4}(\bar x_{1,2},\vec b_{1,2,3,4})&\simeq& \int dz_{1,2,3}^+e^{i\bar x_1 P_2^-(z_1^+-z_4^+)}e^{ix_2 P_2^-(z_2^+-z_3^+)}
\\&&\nn 
\times \langle P_2|
\bar T\{\bar \xi^{b_4}_{\bar n,j_4}(z_4)\bar \xi^{b_3}_{\bar n,j_3}(z_3)\} T\{\xi^{a_2}_{\bar n,i_2}(z_2)\xi^{a_1}_{\bar n,i_1}(z_1)\}
|P_2\rangle\Big|_{z^-_i=0}.
\end{eqnarray}
The visual representation of the terms in cross-section (\ref{DPS_cross_section}) is given in fig.\ref{fig:crosssection_terms}, it also illustrates the ``arrow'' notation for soft factors.

The following steps of classification consists in the Fiertz decomposition of spinor and color structures. As a result of this procedure one gets a large set of various  DPDs with different polarization properties \cite{Diehl:2011yj,Manohar:2012jr,Kasemets:2012pr}. However, the details of Lorentz structure are inessential for the study of soft factor, while the color structure should be considered in details.

In fact, the factorization theorems (\ref{DPS_cross_section},\ref{DPS_cross_section_int}) are not complete, in the sense that they consist of individually singular objects (DPDs and soft factors). They suffer from rapidity divergences, and are not entirely defined. Moreover, the soft factors mix the rapidity divergences related to different sectors of integration. The standard procedure implies that a soft factor can be presented as a product of factors with rapidity divergences from different momentum sectors. Then combining these factors  with appropriate parton distributions one defines an "individual" parton distribution, which are finite and can be used in the phenomenology. Generally, it is unclear (although always implied) whenever it is possible or not to perform the rapidity-factorization procedure and define non-singular DPDs. In the section.\ref{sec:DPD_soft} we demonstrate that such procedure can be done at least at NNLO.

\subsection{Color decomposition}
\label{sec:color_dec}

The color structure of factorized expressions (\ref{DPS_cross_section},\ref{DPS_cross_section_int}) is rather cumbersome. The notation introduced in (\ref{DPS_cross_section}-\ref{DPD_TMD_def3}) specially visualizes the color flow. The indices $a$ and $b$ denote the color adjusted to the antiquark and quark respectively. The subindex of color index designates the position of field in transverse plane (see fig.\ref{fig:crosssection_terms}). In this way, the $SU(N_c)$ gauge transformation transforms DPD such that it is left with respect to indices $a$ and right with respect to indices $b$. For example,
\begin{eqnarray}
F^{b_1b_2a_3a_4}_{qq,\{ij\}}\to U_{b_1b_1'}(\vec b_1)U_{b_2b_2'}(\vec b_2)F^{b'_1b'_2a'_3a'_4}_{qq,\{ij\}}U^\dagger_{a_3'a_3}(\vec b_3)U^\dagger_{a_4'a_4}(\vec b_4),
\end{eqnarray}
where all matrices $U$ are located at light-cone infinities. Consequently, the soft factor transforms in conjugated way by eight matrices $U$. 

In a non-singular gauge the transformation at light-cone infinites can be reduced to unity. In this way, DPDs and soft factors are gauge invariant objects independently.  However, there is the global $SU(N_c)$ rotation of quarks that still transform DPDs. Since global $SU(N_c)$ is a symmetry of QCD, the DPD matrix elements select only the singlet contributions. There are two singlets in $\mathbf{\bar 3\otimes \bar 3\otimes 3 \otimes 3}$ that can be extracted as following
\begin{eqnarray}
F_{q q}^{b_1b_2a_3a_4}&=&\frac{\delta_{b_1a_4}\delta_{b_2a_3}}{N_c^2}F^{\bm 1}_{q q}+\frac{2t^A_{b_1a_4}t^A_{b_2a_3}}{N_c\sqrt{N_c^2-1}}F^{\bm 8}_{q q},
\\
F_{q\bar q}^{b_1a_2b_3a_4}&=&\frac{\delta_{b_1a_4}\delta_{b_3a_2}}{N_c^2}F^{\bm 1}_{q\bar q}+\frac{2t^A_{b_1a_4}t^A_{b_3a_2}}{N_c\sqrt{N_c^2-1}}F^{\bm 8}_{q\bar q},
\\
\mathcal{I}_{\bar q q}^{b_1a_2a_3b_4}&=&\frac{\delta_{b_1a_3}\delta_{b_4a_2}}{N_c^2}\mathcal{I}^{\bm 1}_{\bar q q}+\frac{2t^A_{b_1a_3}t^A_{b_4a_2}}{N_c\sqrt{N_c^2-1}}\mathcal{I}^{\bm 8}_{\bar q q},
\end{eqnarray}
here we use the normalization for singlet parts suggested in \cite{Diehl:2011yj}, which is different from the normalization used in \cite{Manohar:2012jr}. The singlet parts of conjugated distributions are defined in similar manner. Substituting these expressions into (\ref{DPS_cross_section},\ref{DPS_cross_section_int}) we obtain
\begin{eqnarray}\label{DPS_cross_section_singlet}
\frac{d\sigma}{d\vec X}\Big|_{\text{DPS}}&=&d\hat \sigma_{\{ij\}}\int d^2\vec z_{1,2,3} e^{-i\vec{q}_1\cdot(\vec{z}_1-\vec{z}_4)}e^{-i\vec{q}_2\cdot(\vec{z}_2-\vec{z}_3)}\frac{1}{N_c^2}\Bigg[
\\\nn && F_{qq,\{ij\}}S_{\uparrow\uparrow\downarrow\downarrow}\bar F_{\bar q\bar q,\{ij\}}+
F_{\bar q\bar q,\{ij\}}S_{\downarrow\downarrow\uparrow\uparrow}\bar F_{qq,\{ij\}}
\\\nn&&
F_{q\bar q,\{ij\}}S_{\uparrow\downarrow\uparrow\downarrow}\bar F_{\bar q q,\{ij\}}+
F_{\bar q q,\{ij\}}S_{\downarrow\uparrow\downarrow\uparrow}\bar F_{q\bar q,\{ij\}}
\\\nn&&
\mathcal{I}_{\bar qq,\{ij\}}S_{\uparrow\downarrow\downarrow\uparrow}\bar{\mathcal{I}}_{q\bar q,\{ij\}}+
\mathcal{I}_{q\bar q,\{ij\}}S_{\downarrow\uparrow\uparrow\downarrow}\bar{\mathcal{I}}_{\bar qq,\{ij\}}
\Bigg],
\end{eqnarray}
where the DPDs $F$ and $\mathcal{I}$ are 2-component vectors $F=\{F^{\bm 1},F^{\bm 8}\}$, and soft factors are $2\times2$-matrices, which explicit form we present later. The notation (\ref{DPS_cross_section_singlet}) implies the presentation of $\bar F$ as a ``column'', while $F$ as a ``row''. This defines the order of matrix multiplication in the following sections.

\begin{figure}[t]
\centering
\includegraphics[width=0.95\textwidth]{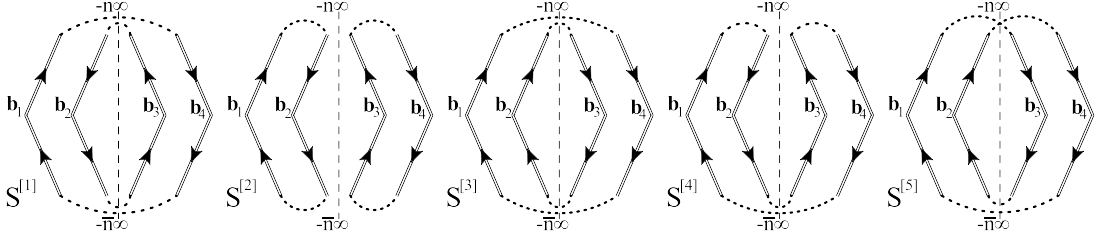}
\caption{Graphical representation of independent components contributing to DPS soft factor (\ref{def:S1}-\ref{def:S5}). Lines with arrows represent Wilson lines. The dashed lines show the index conraction.} \label{fig:DPD_components}
\end{figure}

Before we proceed to the definition of soft-factor matrices, let us discuss the symmetries of components. The obvious symmetry of a soft factor is the exchange of $\Lambda$'s order under the sign of T-ordering. It implies
\begin{eqnarray}\nn
S^{a_1b_1,a_2b_2,a_3b_3,a_4b_4}_{\bullet_1,\bullet_2,\bullet_3,\bullet_4}(\vec b_1,\vec b_2,\vec b_3,\vec b_4)&=&
S^{a_2b_2,a_1b_1,a_3b_3,a_4b_4}_{\bullet_2,\bullet_1,\bullet_3,\bullet_4}(\vec b_2,\vec b_1,\vec b_3,\vec b_4)
\\ \label{SF:sym1}
&=&
S^{a_1b_1,a_2b_2,a_4b_4,a_3b_3}_{\bullet_1,\bullet_2,\bullet_4,\bullet_3}(\vec b_1,\vec b_2,\vec b_4,\vec b_3).
\end{eqnarray}
As the consequence of the Lorentz invariance, the directions $n$ and $\bar n$ can be exchanged within the soft factor independently of the rest expression. Therefore, a soft factor is equal the soft factor with all arrows turned upside-down,
\begin{eqnarray}\label{SF:sym2}
S^{\{ab\}}_{\bullet_1,\bullet_2,\bullet_3,\bullet_4}(\vec b_{1},\vec b_2,\vec b_3,\vec b_4)=S^{\{ab\}}_{\bar \bullet_1,\bar \bullet_2,\bar \bullet_3,\bar \bullet_4}(\vec b_{1},\vec b_2,\vec b_3,\vec b_4),
\end{eqnarray}
where $\bar \bullet=\downarrow (\uparrow)$ for $\bullet=\uparrow(\downarrow)$. Due to these symmetries the soft factors are related to each other. There are only two independent matrices $S_{\uparrow\downarrow\uparrow\downarrow}$ and $S_{\uparrow\uparrow\downarrow\downarrow}$ (as discussed later these two soft factors are also related by (\ref{sf_relations5})). The rest soft factors are expressed as
\begin{eqnarray}\label{sf_relations1}
S_{\downarrow\uparrow\downarrow\uparrow}(\vec b_{1},\vec b_2,\vec b_3,\vec b_4)&=&
S_{\uparrow\downarrow\uparrow\downarrow}(\vec b_{1},\vec b_2,\vec b_3,\vec b_4),
\\\label{sf_relations2}
S_{\downarrow\downarrow\uparrow\uparrow}(\vec b_{1},\vec b_2,\vec b_3,\vec b_4)&=&
S_{\uparrow\uparrow\downarrow\downarrow}(\vec b_{1},\vec b_2,\vec b_3,\vec b_4),
\\\label{sf_relations3}
S_{\uparrow\downarrow\downarrow\uparrow}(\vec b_{1},\vec b_2,\vec b_3,\vec b_4)&=&
S_{\uparrow\downarrow\uparrow\downarrow}(\vec b_{1},\vec b_2,\vec b_4,\vec b_3),
\\\label{sf_relations4}
S_{\downarrow\uparrow\uparrow\downarrow}(\vec b_{1},\vec b_2,\vec b_3,\vec b_4)&=&
S_{\uparrow\downarrow\uparrow\downarrow}(\vec b_{2},\vec b_1,\vec b_3,\vec b_4).
\end{eqnarray}

The soft factor matrix has four components, the Wilson lines of which are connected either by $\delta$'s, or by $SU(N_c)$ generators. In turn, the product of generators $t^At^A$ can be expressed as products of $\delta$'s using Fiertz identities. For practical reasons, it is convenient to consider the products of Wilson lines contracted by $\delta$'s only. There are five independent structures that can appear
\begin{eqnarray}\label{def:S1}
S^{[1]}(\vec b_1,\vec b_2,\vec b_3,\vec b_4)&=&\frac{1}{N_c^2}S_{\uparrow\downarrow\uparrow\downarrow}^{ab,cd,ba,dc}(\vec b_1,\vec b_2,\vec b_3,\vec b_4),
\\
\label{def:S2}
S^{[2]}(\vec b_1,\vec b_2,\vec b_3,\vec b_4)&=&\frac{1}{N_c^2}S_{\uparrow\downarrow\uparrow\downarrow}^{ab,ba,cd,dc}(\vec b_1,\vec b_2,\vec b_3,\vec b_4),
\\
\label{def:S3}
S^{[3]}(\vec b_1,\vec b_2,\vec b_3,\vec b_4)&=&\frac{1}{N_c^2}S_{\uparrow\uparrow\downarrow\downarrow}^{ab,cd,dc,ba}(\vec b_1,\vec b_2,\vec b_3,\vec b_4),
\\
\label{def:S4}
S^{[4]}(\vec b_1,\vec b_2,\vec b_3,\vec b_4)&=&\frac{1}{N_c}S_{\uparrow\downarrow\uparrow\downarrow}^{ab,ca,dc,bd}(\vec b_1,\vec b_2,\vec b_3,\vec b_4),
\\
\label{def:S5}
S^{[5]}(\vec b_1,\vec b_2,\vec b_3,\vec b_4)&=&\frac{1}{N_c}S_{\uparrow\downarrow\downarrow\uparrow}^{ab,cd,da,bc}(\vec b_1,\vec b_2,\vec b_3,\vec b_4).
\end{eqnarray}
A visual representation of these soft-factors is given in fig.\ref{fig:DPD_components}. The normalization is chosen such that at the leading perturbative order all soft factors are unity,
\begin{eqnarray}
S^{[i]}(\vec b_1,\vec b_2,\vec b_3,\vec b_4)=1+\mathcal{O}(a_s).
\end{eqnarray}
The topology of the components is different, so the components $S^{[1,2,3]}$ form two Wilson loops, while $S^{[4,5]}$ are single Wilson loops.

Further simplification of the structure can be made using features of the soft factor geometry special for the Drell-Yan process. The Wilson lines in the matrix element (\ref{SF_gen_def}) are all positioned on the past light cone. Therefore, the distance between any two fields within (\ref{SF_gen_def}) is space-like (or light-like if fields belong to the same Wilson line). It allows to rewrite the T-ordered product of Wilson lines as a usual product of Wilson lines, using the micro-causality relation. However, it is more convenient to organize Wilson lines as a single T-ordered product. We have
\begin{eqnarray}\label{SF_gen_def:Tordered}
S^{\{ab\}}_{\bullet_1\bullet_2\bullet_3\bullet_4}(\vec b_1,\vec b_2,\vec b_3,\vec b_4)=
\langle 0|
T\{\Lambda_{\bullet_4}^{a_4b_4}(\vec b_4)\Lambda_{\bullet_3}^{a_3b_3}(\vec b_3)\Lambda_{\bullet_2}^{a_2b_2}(\vec b_2)\Lambda_{\bullet_1}^{a_1b_1}(\vec b_1)\}
|0\rangle.
\end{eqnarray}
Such presentation is distinctive feature of Drell-Yan kinematic, and is not possible for, say, double semi-inclusive deep-inelastic scattering (SIDIS).

The representation (\ref{SF_gen_def:Tordered}) suggests higher symmetry of soft factor. Namely, the arguments can be freely exchanged preserving the topology of color-connection. Therefore, we need to conciser only soft-factors of different topology: a single Wilson-loop (we choose $S^{[4]}$), and a double Wilson-loop (we choose $S^{[1]}$). The rest are related to the chosen in the following way
\begin{eqnarray}\label{SF:2->1}
S^{[2]}(\vec b_1,\vec b_2,\vec b_3,\vec b_4)&=&S^{[1]}(\vec b_1,\vec b_4,\vec b_3,\vec b_2),
\\\label{SF:3->1}
S^{[3]}(\vec b_1,\vec b_2,\vec b_3,\vec b_4)&=&S^{[1]}(\vec b_1,\vec b_3,\vec b_2,\vec b_4),
\\\label{SF:5->4}
S^{[5]}(\vec b_1,\vec b_2,\vec b_3,\vec b_4)&=&S^{[4]}(\vec b_1,\vec b_3,\vec b_2,\vec b_4).
\end{eqnarray}
We also find that the soft factor $S_{\uparrow\downarrow\downarrow\uparrow}$ can be expressed via $S_{\uparrow\downarrow\uparrow\downarrow}$ as
\begin{eqnarray}\label{sf_relations5}
S_{\uparrow\uparrow\downarrow\downarrow}(\vec b_1,\vec b_2,\vec b_3,\vec b_4)=S_{\uparrow\downarrow\uparrow\downarrow}(\vec b_1,\vec b_3,\vec b_2,\vec b_4).
\end{eqnarray}
Therefore, we can consider only the case of $S_{\uparrow\downarrow\uparrow\downarrow}$, while the results for other channels can be obtained by permuting vectors $\vec b_i$.

Using the symmetries of soft factor (\ref{SF:sym1}-\ref{SF:sym2}) and the notation for independent components (\ref{def:S1},\ref{def:S4}), we present the $2\times2$-matrix $S_{\uparrow\downarrow\uparrow\downarrow}$ in the form
\begin{eqnarray}\label{Sudud}
S_{\uparrow\downarrow\uparrow\downarrow}=\left(
\begin{array}{cc}
S^{[1]}(\vec b_{1,3,4,2}) &\Ds{\frac{S^{[4]}(\vec b_{1,2,3,4})-S^{[1]}(\vec b_{1,2,3,4})}{\sqrt{N_c^2-1}}}
\\
\Ds{\frac{S^{[4]}(\vec b_{1,2,3,4})-S^{[1]}(\vec b_{1,2,3,4})}{\sqrt{N_c^2-1}}} & \frac{N_c^2 S^{[1]}(\vec b_{1,4,3,2})+S^{[1]}(\vec b_{1,2,3,4})-2S^{[4]}(\vec b_{1,2,3,4})}{N_c^2-1}
\end{array}
\right).
\end{eqnarray}
Here, for compactness we use the shorthand notation for the argument $S(\vec b_{i,j,k,l})=S(\vec b_i,\vec b_j,\vec b_k,\vec b_l)$. The rest of the soft factor matrices can be obtained via (\ref{sf_relations1}-\ref{sf_relations4},\ref{sf_relations5}). At the leading order of perturbation theory  soft factor matrices reduces to identity matrices
\begin{eqnarray}
S_{\bullet_1\bullet_2\bullet_3\bullet_4}(\vec b_1,\vec b_2,\vec b_3,\vec b_4)=\left(\begin{array}{cc}1 &0 \\ 0 &1
\end{array}\right)+\mathcal{O}(a_s).
\end{eqnarray}

\section{Evaluation of soft factors}
\label{sec:sec3}

\subsection{Generating function for web diagrams}
\label{sec:generating_function}

The straightforward evaluation of functions $S^{[i]}$ requires a calculation of many diagrams, most of which are equivalent under permutation of parameters and change of color factors. Such a consideration would be very inefficient and contains many potential places for a mistake. A more effective approach is to evaluate the generating function for web diagrams, which is common for all soft factors, and project out the appropriate soft factor. The theoretical description of the approach can be found in \cite{Vladimirov:2015fea,Vladimirov:2014wga}. In this section we describe only the basics of generating function approach needed for this particular calculation.

The generating function approach is based on the well-known fact that the perturbative series for vacuum average of some operator sources is an exponent of the connected diagrams. This property immediately leads to exponentiation theorem for Wilson lines for Abelian gauge theories \cite{Vladimirov:2014wga}. For a non-Abelian gauge theories one has an additional difficulty coming from the necessity to disentangle the color structure. The disentangling can be done in the general form \cite{Vladimirov:2015fea}. In this way, one sees that significant part of diagrams that appear in the usual perturbation expansion (as well as, in the classical Wilson loop exponentiation diagrammatic \cite{Gatheral:1983cz,Frenkel:1984pz}) are composed from the smaller-loop diagrams. 

The power of the generating functions approach is that evaluated ones the generating function can be easily used to obtain the perturbative expression for any color topology. Therefore, the generating function that we present later can be used to obtain all DPS soft factors (\ref{def:S1}-\ref{def:S5}), as well as, TMD soft factor and soft factors for multi parton scattering with six, and more operators $\Lambda$. Moreover, the approach allows one to consider the exponentiated expression directly in the matrix form.

The starting point of the construction is to carry out the color structure of the Wilson line. The effective way to do so is to introduce the "scalar-reduction" of a Wilson line connecting points $x$ and $y$ \cite{Vladimirov:2015fea}
\begin{eqnarray}
W_n[x,y]=e^{t^A\frac{\partial}{\partial \theta^A}}e^{\theta^A V^n_A[x,y]}\Big|_{\theta=0},
\end{eqnarray}
where $t^A$ is a gauge-group generators, $\theta^A$ are c-number variables, and $V$ is functional of gauge fields. The particular form of $V$ needed for our calculation is given in (\ref{def:op_V}), while the general form can be found in \cite{Vladimirov:2015fea,Vladimirov:2014wga}. In this expression the matrix structure is carried by the first exponent in the product. The first exponent does not contain any fields and thus, does not participate in the function integration. The Wilson line flowing in the opposite direction can be presented in the form
\begin{eqnarray}
W_n[y,x]=\(W_n[x,y]\)^\dagger=e^{-t^A\frac{\partial}{\partial \theta^A}}e^{\theta^A V^n_A[x,y]}\Big|_{\theta=0},
\end{eqnarray}
where we have used that operator $V$ is anti-hermitian $V^\dagger=-V$.

Within the considered task, we have a simple geometry of Wilson lines. All of them are straight (along $n$ or $\bar n$), and continue from $\vec b_i$ to  infinity (or in opposite direction). Let us enumerate these segments by number $j$ (for DPS soft factor $j=1,..,8$). The $j$'th Wilson line can be presented in the form
\begin{eqnarray}\label{scalar_reduction_of_WL}
W^{r_j}_{v_j}[\vec b_j,v_j\infty]=e^{r_jt^A\frac{\partial}{\partial \theta_j^A}}e^{\theta_j^A V^j_A(v_j,\vec b_j)}\Big|_{\theta=0},
\end{eqnarray}
where $v_j=$($n$ or $\bar n$), the variable $r_j$ denoted the direction $r_j=1(-1)$ if the color flow to (from) light-cone infinity. The operator $V$, which discribes a half infinite Wilson-line is given by \cite{Vladimirov:2015fea}
\begin{eqnarray}\label{def:op_V}
V^{i}_A(v_i,\vec b_i)&=&-i gv^\mu_i\int_0^{\infty} d\sigma  A_\mu^A(\vec b_{i}-v_i\sigma)
\\\nn &&+i\frac{g^2}{2}f^{ABC}v^\mu_iv^\nu_i\int_0^{\infty} d\sigma\int_0^{\sigma} d\tau
A^B_\mu(\vec b_{i}-v_i\sigma)A^C_\nu(\vec b_{i}-v_i\tau)+\mathcal{O}(g^3).
\end{eqnarray}
In the following we omit the arguments of operator $V$ for brevity. The functional $V$ is an infinite series of path-ordered gauge field commutators. Here we truncate the series at $g^2$ order which is enough for NNLO calculation. The functionals $V$ have many nice properties within the perturbation theory, especially, for light-like paths (see \cite{Vladimirov:2015fea}). We use these properties for extra check of loop-calculation.

Using the expression (\ref{scalar_reduction_of_WL}) we obtain the following expression for a generic soft factor
\begin{eqnarray}\label{formula1}
&&\langle 0|T\{ \Lambda_\bullet^{a_1b_1}(\vec b_1)...\Lambda_{\bullet}^{a_nb_n}(\vec b_n)\}|0\rangle 
\\\nn
&&\qquad\qquad=\(e^{-t^A\frac{\partial}{\partial \theta_1^A}}e^{t^B\frac{\partial}{\partial \theta_2^B}}\)^{a_1b_1}...\(e^{-t^A\frac{\partial}{\partial \theta_{2n-1}^A}}e^{t^B\frac{\partial}{\partial \theta_{2n}^B}}\)^{a_nb_n} 
\langle 0| e^{\sum_j \theta^A_j V_A^j} |0\rangle\Big|_{\theta=0},
\end{eqnarray}
where we explicitly write all color indices. The right-hand-side of expression (\ref{formula1}) is a product of the color projector and the colorless matrix element. 

In the case of a Wilson loop it is natural to rename the indices $j$ such that they are ordered along the Wilson loop. The color projection operator in this case has especially simple form of P-ordered exponent of derivative operators, a discrete analog of Wilson line,
\begin{eqnarray}\label{def:proj_1WL}
e^{r_1t^A\frac{\partial}{\partial \theta_1^A}}...e^{r_nt^A\frac{\partial}{\partial \theta_n^A}}  
\xrightarrow{\text{Wilson loop}} \Tr P\exp\(\sum_j r_jt^A\frac{\partial}{\partial \theta_j^A}\),
\end{eqnarray}
where P-ordering is made with respect to index $j$. In the case of double Wilson loop the projector is product of two traces
\begin{eqnarray}\label{def:proj_2WL}
e^{r_1t^A\frac{\partial}{\partial \theta_1^A}}...e^{r_nt^A\frac{\partial}{\partial \theta_n^A}}\xrightarrow{\text{double W.l.}} \Tr P\exp\(\sum_{j\in L1} r_jt^A\frac{\partial}{\partial \theta_j^A}\)\Tr P\exp\(\sum_{j\in L2} r_jt^A\frac{\partial}{\partial \theta_j^A}\),
\end{eqnarray}
where subsets $L1$ and $L2$ of indices $j$ are separately ordered along the first and the second loops.

The matrix element on the right-hand-side of (\ref{formula1}) can be presented as
\begin{eqnarray}
\langle 0| e^{\sum_j \theta^A_j V_A^j} |0\rangle=e^{W[\theta]},
\end{eqnarray}
where $W$ is the generating function for web diagrams. It is given by the sum of amplitudes
\begin{eqnarray}
W[\theta]=\frac{1}{2}\sum_{j_1j_2}\theta_{j_1}^A\theta_{j_2}^B \langle V_A^{j_1}V_B^{j_2} \rangle+
\frac{1}{3!}\sum_{j_1j_2j_3}\theta_{j_1}^A\theta_{j_2}^B\theta^C_{j_3} \langle V_A^{j_1}V_B^{j_2}V_C^{j_3} \rangle+...~,
\end{eqnarray}
where $\langle O \rangle$ is the connected part of the vacuum expectation value of the operator $O$.  We have dropped the term linear in $\theta$ since it does not contribute in light-like kinematics. The dots denote the matrix elements with higher number of $V$. The first dropped contribution is $\sim V^4$, which is of order $g^6$ and hence NNNLO. 

To obtain the expression for a soft factor we need to evaluate the correlator of two and three operators $V$ at $g^4$ order. The different soft factors are obtained by application of different the color-projection operator.

\subsection{Generating function at order $g^2$}
\label{sec:1loop}

\begin{figure}[t]
\centering
\includegraphics[width=0.185\textwidth]{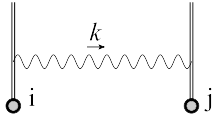}
\caption{Diagram contributing to the leading order of generating function.} \label{fig:1loop}
\end{figure}

In this section we discuss the evaluation of generating function at order $g^2$ in details. The calculation is almost trivial but we use it for the clarification of notation. Since $V\sim g$, only the correlator of two operators $V$ contributes at $g^2$ order. It is given by a single shown in fig.\ref{fig:1loop}.

The soft factor diagrams contains rapidity divergences. To regularize them we use the $\delta$-regularization scheme as it is defined in \cite{Echevarria:2015byo}. Roughly speaking, the $\delta$-regularization consists in the multiplication of every gluon field $A(\sigma n)$ within a Wilson line by the exponent factor $e^{-\delta \sigma}$.  So the gluon field is forced to zero at light-cone infinity. The positive infinitesimal parameters $\delta$ are different for $n$- and $\bar n-$ infinities, $\delta^+$ and $\delta^-$ respectively. 

The Feynman rule for a single-gluon interaction with operator $V_i^A$ is given by
\begin{eqnarray}
\frac{\delta}{\delta A_\mu^A(k)}V_i^B \theta_i^B\Big|_{A_\mu=0} = \frac{g v_i^\mu \theta_i^A e^{i \vec k \cdot \vec b_i}}{k_i+i\delta_i},
\end{eqnarray}
where $k_i=(v_i\cdot k)$, $\delta_i=\delta^+$($\delta^-$) for $v_i=n(\bar n)$ and the momentum is incoming to vertex. The expression for diagram shown in fig.\ref{fig:1loop} is
\begin{eqnarray}\label{1loop:A}
\text{Diag}_A&=&\theta^A_i\theta^A_j v_{ij} g^2 \int \frac{d^dk}{(2\pi)^d}\frac{-i e^{i \vec k \vec b_{ij}}}{(k_i+i\delta_i)(-k_j+i\delta_j)(k^2+i0)},
\end{eqnarray}
where $v_{ij}=(v_i\cdot v_j)$, $\vec b_{ij}=\vec b_i-\vec b_j$.

The following observations are useful also for many two-loop diagrams.
\begin{itemize}
\item The loop integrals are invariant under separate change of sign for transverse and light-like components. Therefore, the (overall) sign in the exponent(s) is irrelevant.
\item The diagram is proportional to $v_{ij}$, which is zero if both $v_i$ and $v_j$ along the same direction. Therefore, under the sign of integral we can set $i=n$ and $j=\bar n$ without loss of generality. In this way, the loop-integral is independent on the vectors $v_i$ and $v_j$, while this dependence is given solely by the prefactor $v_{ij}$.
\end{itemize}
Using these observations we combine terms of the generating function at $g^2$ into very simple form
\begin{eqnarray}\label{W_1loop}
W[\theta]=\frac{a_s}{2}\sum_{ij}\theta_i^A\theta_j^A v_{ij} K^{(0)}(\vec b_{ij}),
\end{eqnarray}
where $a_s=g^2/(4\pi)^2$ is QCD perturbative parameter. The explicit expression for $K^{(0)}$ is presented in (\ref{app:int_K}). 

The transverse distances $\vec b_{ij}$ within the perturbative expansion of soft factors are formally unrestricted. However, for the large values of $\vec b_{ij}$ the logarithm contributions in (\ref{W_1loop}) became large, and violate the convergence of the perturbative series. Therefore, practically the values of $\vec b_{ij}$ should be restricted as $a_s(\mu)\ln(\vec b^2_{ij}\mu^2)<1$. 

\subsection{Generating function at order $g^4$}
\label{sec:2loop}

\begin{figure}[t]
\centering
\includegraphics[width=0.95\textwidth]{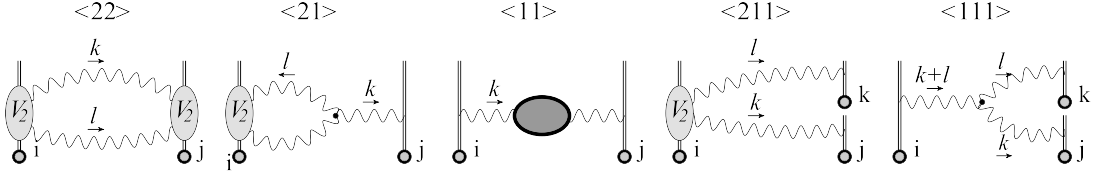}
\caption{Diagrams contributing to the generating function at next-to-leading order.} \label{fig:2loop}
\end{figure}

The diagrams contributing to the generating function at NNLO are presented in fig.\ref{fig:2loop}. In many aspects the calculation repeats the one loop calculation and details are presented in appendix \ref{app:calc}. To evaluate the diagrams one needs the Feynman rules for the radiation of two gluon from the effective vertex. It is
\begin{eqnarray}\label{V2_FeynRules}
\frac{\delta^2}{\delta A_{\mu_1}^{B_1}(k_1)\delta A_{\mu_2}^{B_2}(k_2)}V^A_i\theta^A_i\Big|_{A_\mu=0}\!\!\! = i f^{AB_1B_2} \theta^A_i\frac{g^2}{2}\(\frac{1}{k_{2i}+i\delta_i}-\frac{1}{k_{1i}+i\delta_i}\)\frac{v_i^{\mu_1}v_i^{\mu_2}e^{i(\vec k_1+\vec k_2)\vec b_i}}{k_{1i}+k_{2i}+2i\delta_i},
\end{eqnarray}
where all momenta are incoming to vertex.

Let us write the general form for the generating function at NNLO. It can be found from the symmetries of matrix elements. First, due to the global $SU(N_c)$ symmetry a matrix element $\langle V^A_i...V^B_j\rangle$ is proportional to the invariant $SU(N_c)$ tensors only. Second, a matrix element $\langle V^A_i...V^B_j\rangle$ is invariant under permutation of operators $V$. And finally, due to the Lorentz invariance and power counting the functions $W$ can have the scalar products $v_{ij}$ only as a prefactor. Combining together these observations the generating function at NNLO can be parametrized in the terms of two functions
\begin{eqnarray}\label{gen_func:gen}
W[\theta]=\frac{a_s}{2}\!\!\sum_{j_1,j_2}\theta^A_{j_1}\theta^A_{j_2}W_{j_1,j_2} +\frac{a_s^2}{3!}\!\!\sum_{j_1j_2j_3}\!\!if^{ABC} \theta^A_{j_1}\theta^B_{j_2}\theta^C_{j_3}W_{j_1j_2j_3}+\mathcal{O}(a_s^3),
\end{eqnarray}
where 
\begin{eqnarray}\label{def:Wij}
W_{ij}&=&v_{ij}W[\vec b_{ij}],
\\\label{def:Wijk}
W_{ijk}&=&v_{ij}v_{ik}W[\vec b_{ij},\vec b_{ik},\vec b_{jk}].
\end{eqnarray}
We also confirm this form by direct calculation presented in appendix \ref{app:calc}. Note, that at NNNLO the structure of the generating function is reacher. It contains a term proportional to $d^{ABC}$ and various terms of the form $f^{AB\alpha}f^{\alpha CD}$ and $f^{AB\alpha}d^{\alpha CD}$.

The evaluation of diagrams contributing to $W[\vec b]$ is nearly in one-to-one correspondence with the evaluation of similar diagrams in the case of TMD soft-factor made in \cite{Echevarria:2015byo}. The explicit expression for functions $W$ within the $\delta$-regularization are
\begin{eqnarray}\nn
W[\vec b]&=&K^{(0)}(\vec b)+a_s\Bigg[\frac{C_A}{2}\(
I_A''(\vec b)-\frac{(K^{(0)}(\vec b))^2}{4}+2
(2I_{C1}(\vec b)+I_{C2}(\vec b))-\frac{4}{\epsilon}K^{(0)}(\vec b)\)
\\&&\label{gen_func:2V} -
\(C_A(5-3\epsilon)-2N_f(1-\epsilon)\)\frac{2\Gamma(2-\epsilon)\Gamma(-\epsilon)\Gamma(1+\epsilon)}{\Gamma(4-2\epsilon)} K^{(\epsilon)}(\vec b)
\\&&\nn
\qquad\qquad\qquad\qquad\qquad\qquad\qquad
-\frac{1}{\epsilon}\(\frac{5}{3}C_A-\frac{2}{3}N_f\)K^{(0)}(\vec b)
\Bigg]+\mathcal{O}(a_s^2),
\\\label{gen_func:3V}
W[\vec b_1,\vec b_2,\vec b_3]&=&J(\vec b_1,\vec b_2)+2R(\vec b_2,\vec b_3)+\mathcal{O}(a_s),
\end{eqnarray}
where base loop-integrals are given in appendix \ref{app:loop_integrals}. One can see that the complete NNLO expression for generating function contains only five basis integrals $K^{(a)}$, $I''_A$, $I_{C1}$, $I_{C2}$, $J$ and $R$, that are given in (\ref{app:int_K},\ref{app:Int_IA},\ref{app:Int_IC1},\ref{app:Int_IC2},\ref{app:Int_J},\ref{app:Int_R2}). These loop integrals can be compared and agree the loop integrals evaluated in \cite{Echevarria:2015byo}.

\subsection{Action of projection operator}
\label{sec:action_of_proj}

As it was shown any soft factor with topology of a single Wilson loop can be obtained by the action of operator (\ref{def:proj_1WL}) on the generating function (\ref{gen_func:gen}). The result of the action reads
\begin{eqnarray}\label{single_loop_SF}
S_{L}&=&\frac{1}{N_c}\Tr P e^{\sum_j r_jt^A\frac{\partial}{\partial \theta^A_j}}e^{W[\theta]}\Big|_{\theta=0}=\exp\Bigg\{a_s C_F \sum_{j_1<j_2}r_{j_1}r_{j_2}W_{j_1j_2}
\\\nn&& \qquad\qquad +a_s^2 \(-\frac{C_FC_A}{2}\)\sum_j\Bigg[\frac{1}{4}W^2_{j_1j_2}+r_{j_1}r_{j_2}r_{j_3}W_{\{j_1j_2j_3\}}+
\\\nn &&\qquad\qquad +\frac{1}{2}\(r_{j_1}r_{j_3}W_{j_1j_2}W_{j_2j_3}+r_{j_2}r_{j_3}W_{j_1j_2}W_{j_1j_3}+r_{j_1}r_{j_2}W_{j_1j_3}W_{j_2j_3}\)
\\\nn &&\qquad\qquad +r_{j_1}r_{j_2}r_{j_3}r_{j_4}W_{j_1j_3}W_{j_2j_4}\Bigg]+\mathcal{O}(a_s^3)\Bigg\},
\end{eqnarray}
where sums over $j$ are strictly ordered, i.e. $j_1<j_2<...<j_n$. Here and later, we use notation $C_F=(N_c^2-1)/2N_c$ and $C_A=N_c$ where it is convenient. To derive the expression (\ref{single_loop_SF}) we have used that $W_{ij}=W_{ji}$ and that $r_j^2=1$. The curly brackets on the indices of the third term denote the anti-symmetrization over permutation of indices (with $1/3!$ prefactor). One can see that the color factors which appear in the exponent corresponds to the color-connected parts of diagrams, in accordance of the exponentiation theorem for a Wilson loop \cite{Gatheral:1983cz,Frenkel:1984pz}.

The topology of the double Wilson-loop (composed of loops $L1$ and $L2$) is described by the projection operator (\ref{def:proj_2WL}). Applying (\ref{def:proj_2WL}) on the generating function (\ref{gen_func:gen}) we obtain
\begin{eqnarray}\label{double_loop_SF}
&&\frac{1}{N_c}\Tr P e^{\sum_{i\in L1} r_it^A\frac{\partial}{\partial \theta^A_i}}\frac{1}{N_c}\Tr P e^{\sum_{j\in L2} r_jt^A\frac{\partial}{\partial \theta^A_j}}e^{W[\theta]}\Big|_{\theta=0}=S_{L1}S_{L2}\times
\\\nn&&\qquad\qquad\exp\Bigg\{a_s^2\(\frac{C_FC_A}{2}-C_F^2\)\sum_{i,j}\Bigg[
\frac{1}{2}W^2_{i_1j_1}+r_{i_1}r_{j_1}r_{j_2}W_{i_1j_1}W_{i_1j_2}+r_{i_1}r_{i_2}r_{j_1}W_{i_1j_1}W_{i_2j_1}
\\\nn &&\qquad\qquad +r_{i_1}r_{i_2}r_{j_1}r_{j_1}(W_{i_1j_1}W_{i_2j_2}+W_{i_1j_2}W_{i_2j_1})\Bigg]+\mathcal{O}(a_s^3)\Bigg\},
\end{eqnarray}
where $S_{L1}$ and $S_{L2}$ are soft factors evaluated on a single loop and defined in  (\ref{single_loop_SF}), and indices $i$ and $j$ belong to loops $L1$ and $L2$, respectively, and are strictly ordered along loops. The last two lines of (\ref{double_loop_SF}) represent the interaction of Wilson loops. The leading order of  between-loops interaction is given by double-gluon exchange, and, thus, is NNLO in coupling constant.

The expression for double Wilson loop topology (\ref{double_loop_SF}) can be easily generalized for the case of arbitrary number of Wilson loops, because at NNLO only two Wilson loops can simultaneously interact with each other. Denoting the interaction between loops $L1$ and $L2$ (given in the last two lines of (\ref{double_loop_SF})) as $S_{L1L2}$ we obtain
\begin{eqnarray}
\frac{1}{N_c}\Tr P e^{\sum_{i\in L1} r_it^A\frac{\partial}{\partial \theta^A_i}}...\frac{1}{N_c}\Tr P e^{\sum_{j\in Ln} r_jt^A\frac{\partial}{\partial \theta^A_j}}e^{W[\theta]}\Big|_{\theta=0}=\prod_n S_{Ln}\prod_{n< k} S_{LnLk}+\mathcal{O}(a_s^3).
\end{eqnarray}

\subsection{TMD soft factor}
\label{sec:TMD_SF}

Before we proceed further it is instructive to calculate the TMD soft factor. Within the $\delta$-regularization the TMD soft factor has been evaluated at NNLO in \cite{Echevarria:2015byo}, and successfully used for description of TMD parton distribtions and TMD fragmentation functions at NNLO in \cite{Echevarria:2015usa,Echevarria:2016scs}. Recently, TMD soft factor has been evaluated at NNLO in the rapidity regularization \cite{Luebbert:2016itl}.

The TMD soft factor (in Drell-Yan kinematics) is given by the following matrix element
\begin{eqnarray}
S^{\text{TMD}}=\langle 0|T\{\Lambda^{ab}_\uparrow (\vec b)\Lambda_\downarrow^{ba}(\vec 0)\}|0\rangle.
\end{eqnarray}
In this case the index $j$ runs from $1$ to $4$ and the parameters of the TMD soft factor are
\begin{eqnarray}
r_j=(-1)^{j+1},\qquad \vec b_j=\{\vec b,\vec 0,\vec 0,\vec b\}.
\end{eqnarray}
Evaluating the expression (\ref{single_loop_SF}) with these parameters we obtain
\begin{eqnarray}\label{TMD_SF}
\ln S^{\text{TMD}}&=&\sigma(\vec b)=2C_Fa_s\(W(\vec b)-W(\vec 0)\)+C_FC_Aa_s^2\(W(\vec b,\vec 0,\vec b)-W(\vec 0,\vec b,\vec b)\)
\\&&\nn\qquad\qquad\qquad-\frac{C_FC_A}{4}a_s^2\(3 W^2(\vec b)-4 W(\vec b)W(\vec 0)+W^2(\vec 0)\)+\mathcal{O}(a_s^3).
\end{eqnarray}
To present (\ref{TMD_SF}) in such a form, we have used that
\begin{eqnarray}\label{relation_for_W}
W(-\vec b)=W(\vec b),\qquad W(-\vec b_1,\vec b_2,\vec b_3)=W(\vec b_1,-\vec b_2,-\vec b_3)=W(\vec b_1,\vec b_2,\vec b_3).
\end{eqnarray}
These relations follow from the expressions of loop integrals (\ref{app:R_sym}). Generally speaking, these symmetries are presented only at NNLO, but they are not crucial for further development and used only for visual simplification of the result. Substituting the explicit expressions for loop integrals into (\ref{TMD_SF}) we have checked that result (\ref{TMD_SF}) coincides with one presented in \cite{Echevarria:2015byo}.

\subsection{Soft factor $S^{[4]}$}
\label{sec:S4}

Let us consider the soft factor $S^{[4]}$. We choose the enumeration of Wilson segments such that it starts at the point indicated in fig.\ref{fig:DPD_components} as $\vec b_1$, and follows the arrows of color flow. Therefore, we have $r_j=(-1)^{j+1}$, and
\begin{eqnarray}
b_j=\{\vec b_1,\vec b_2,\vec b_2,\vec b_3,\vec b_3,\vec b_4,\vec b_4,\vec b_1\}.
\end{eqnarray}
Evaluating the formula (\ref{single_loop_SF}) with these parameters we obtain an expression in terms of functions $W$ (\ref{def:Wij},\ref{def:Wijk}). Considering this expression one can recognize the entries in the form of (\ref{TMD_SF}). It appears that the soft factor $S^{[4]}$ can be conveniently written via the TMD soft factor only. In terms of the function $\sigma$ introduced in (\ref{TMD_SF}) the soft factor $S^{[4]}$ is remarkably simple
\begin{eqnarray}\label{S4}
\ln S^{[4]}&=&\sigma(\vec b_{12})-\sigma(\vec b_{13})+\sigma(\vec b_{14})+\sigma(\vec b_{23})-\sigma(\vec b_{24})+\sigma(\vec b_{34})
\\\nn &&+\frac{C_A}{4C_F}(\sigma(\vec b_{13})-\sigma(\vec b_{14})-\sigma(\vec b_{23})+\sigma(\vec b_{24}))(\sigma(\vec b_{12})-\sigma(\vec b_{13})-\sigma(\vec b_{24})+\sigma(\vec b_{34}))+\mathcal{O}(a_s^3).
\end{eqnarray}
To derive it we have used relations (\ref{relation_for_W}). 

An additional check is granted by the fact that setting two subsequent $\Lambda$'s at the same point we obtain the TMD soft-factors. Namely, 
\begin{eqnarray}
S^{[4]}(\vec b_1,\vec b_2,\vec b_2,\vec b_4)&=&S^{\text{TMD}}(\vec b_{14}),
\\
S^{[4]}(\vec b_1,\vec b_2,\vec b_3,\vec b_1)&=&S^{\text{TMD}}(\vec b_{23}),
\\
S^{[4]}(\vec b_1,\vec b_1,\vec b_3,\vec b_4)&=&S^{\text{TMD}}(\vec b_{34}),
\\
S^{[4]}(\vec b_1,\vec b_2,\vec b_3,\vec b_3)&=&S^{\text{TMD}}(\vec b_{12}).
\end{eqnarray}

One can see that there is no direct three-lines correlations at this order. The multi-line correlation appears only as a product of pairwise interactions (the second line in (\ref{S4})). In fact, this is a general feature of multi-particle soft factors and can be seen on the diagram level. Let us describe this effect at NNLO, although some part of statements can be generalized to arbitrary order.

\begin{figure}[t]
\centering
\includegraphics[width=0.45\textwidth]{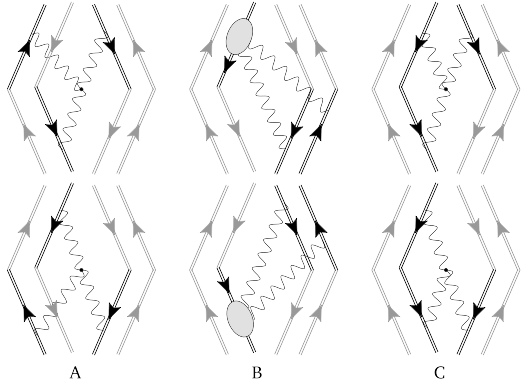}
\caption{Top and bottom diagrams of type A and B have the same expression and color factor but different signs due to the orientation of Wilson lines and, thus, cancel each other. Top and bottom diagrams of type C have same expression, but different signs due to the orientation of Wilson lines, different signs of color factors and, thus, sum together.} \label{fig:Discus}
\end{figure}

At NNLO one has only two topologies of diagrams that connect three Wilson line. They are shown in fig.\ref{fig:Discus}. Every such diagram has a partner that is "reflected" upside-down (compare top and bottom diagrams in fig.\ref{fig:Discus}). The loop-integrals for pairs of such diagrams are the same, due to Lorentz invariance. However, the difference between diagrams can appear because of the different color connection and directions of Wilson lines. It is clear that the "reflected" diagrams necessary have opposite general sign due to the direction of Wilson lines, i.e. the combination $r_{i}r_{j}r_k$ changes sing under the "reflection". If three Wilson lines belong to different $\Lambda$'s (see diagrams A and B in fig.\ref{fig:Discus}) then the "reflected" diagrams have the same color factor. Therefore, the diagrams that connects three different $\Lambda$'s cancel each other. For the diagrams that connect two $\Lambda$'s (see diagrams C in fig.\ref{fig:Discus}) the color factor also (together with the general sign) changes the sign under "reflection", and thus, these diagrams doubled in the final result. Therefore, the diagrams that connect three $\Lambda$'s drop from the soft factors. As consequence DPS soft factor can be expressed in via TMD soft factors only. 

The discussed cancellations are not transparent within a usual diagrammatic. The diagrams of type B in fig.\ref{fig:Discus}  are not presented in the usual diagrammatic. Instead there is a set of diagrams with two gluons connected to the Wilson line in different orders. The corresponding "reflected" diagrams have different loop-integrals, and cannot be directly compared. To perform comparison one should split these diagrams to symmetric (that part reduces to one-loop integrals) and anti-symmetric parts (that part is irreducible). The effective vertex $V_2$ represents the anti-symmetric contribution. This contribution cancels in the sum of mirrored diagrams. The symmetric combination is reducible and reveals after action of projection operator. Therefore, the generating function approach is effective tool for consideration of many Wilson lines configurations.

\subsection{Soft factor $S^{[1]}$}
\label{sec:S1}

Let us consider the soft factor $S^{[1]}$. We start the enumeration for the first loop from the point indicated in fig.\ref{fig:DPD_components} as $\vec b_1$, and for the second loop from the point indicated as $\vec b_3$. Therefore, we have $r_j=(-1)^{j+1}$, and $r_i=(-1)^{i+1}$, and
\begin{eqnarray}
b_i=\{\vec b_1,\vec b_4,\vec b_4,\vec b_1\},\qquad b_j=\{\vec b_3,\vec b_2,\vec b_2,\vec b_3\}.
\end{eqnarray}
Evaluating expression (\ref{double_loop_SF}) with these parameters, with accordance of discussion given in the previous section, we obtain expression in the terms of the TMD soft factor only. Using the function $\sigma$ introduced in (\ref{TMD_SF}) we obtain
\begin{eqnarray}\label{S1}
\ln S^{[1]}&=&\sigma(\vec b_{14})+\sigma(\vec b_{23})+\frac{1}{2}\(\frac{C_A}{2C_F}-1\)(\sigma(\vec b_{12})-\sigma(\vec b_{13})-\sigma(\vec b_{24})+\sigma(\vec b_{34}))^2+\mathcal{O}(a_s^3).
\end{eqnarray}

An additional check is granted by the fact that setting two subsequent $\Lambda$'s at the same point we obtain the TMD soft-factor. Namely, 
\begin{eqnarray}
S^{[1]}(\vec b_1,\vec b_2,\vec b_2,\vec b_4)&=&S^{\text{TMD}}(\vec b_{14}),
\\
S^{[1]}(\vec b_1,\vec b_2,\vec b_3,\vec b_1)&=&S^{\text{TMD}}(\vec b_{23}).
\end{eqnarray}

Substituting the expressions (\ref{S4}) and (\ref{S1}) for components $S^{[1,4]}$ in to the matrix (\ref{Sudud}), we obtain the explicit expression for $S_{\uparrow\downarrow\uparrow\downarrow}$. The rest soft factors are obtained by the permutation of arguments as discussed in the section \ref{sec:color_dec}. We have checked that at NLO these expressions coincides with ones calculated in \cite{Diehl:2011yj}.

\section{DPS soft factors and separation of rapidity divergences}
\label{sec:DPD_soft}
\subsection{Recombination of rapidity divergences}
\label{sec:recombine}

As we have discussed in sec.\ref{sec:factorization}, the factorization formula (\ref{DPS_cross_section_singlet}) is incomplete, in the sense, that the collinear matrix elements $F$ and soft-factor contain rapidity divergences. To complete the factorization formula and to construct a well-defined DPDs one has to disentangle rapidity divergences of $n-$ and $\bar n$-soft Wilson lines of the soft factor. These divergences recombine with the divergences arising in corresponding collinear matrix elements. 

This procedure is very intuitive in the case of TMD factorization, let us briefly remind it. Within $\delta$-regularization, the TMD soft factor (\ref{TMD_SF}) is strictly linear in $\ln(\delta^+\delta^-)$ 
\begin{eqnarray}
\sigma(\vec b)=A(\vec b)\ln\(\frac{\mu^2}{\delta^+\delta^-}\)+B(\vec b).
\end{eqnarray}
The variables $\delta^+$ and $\delta^-$ regularize the rapidity divergences in $n$ and $\bar n$ direction, respectively. The same regulator appears in $n-$ and $\bar n-$ collinear matrix elements, such that the cross-section is finite. Schematically, the TMD cross-section can be written as
\begin{eqnarray}\label{TMD_crossSec_scheme}
d\sigma\sim F^{sub}\(\frac{\delta^+}{P^+}\)\times S\(\frac{\mu^2}{\delta^+\delta^-}\)\times \bar F^{sub}\(\frac{\delta^-}{P^-}\).
\end{eqnarray}
Here the superscript $sub$ denotes the collinear matrix element with subtracted overlap modes (zero-bin subtraction). Due to the linearity of soft factor in $\ln(\delta^+\delta^-)$ one can easily separate divergences in $\delta^+$ from divergences in $\delta^-$,
\begin{eqnarray}\label{TMD_SF_dec}
\sigma(\vec b)=\sigma^+(\vec b)+\sigma^-(\vec b),\qquad \sigma^\pm(\vec b)=\frac{A(\vec b)}{2}\mathbf{l}^\pm+\frac{B(\vec b)}{2},
\end{eqnarray}
where 
$$
\mathbf{l}^\pm=\ln\(\frac{\mu^2}{(\delta^\pm/P^\pm)^2\zeta_\pm}\),
$$
and $\zeta_+\zeta_{-}=(P^+_1P^-_2)^2\sim Q^4$. Combining the parts of the soft factor with collinear matrix elements we obtain a well-defined "individual" TMD \cite{Echevarria:2012js}
\begin{eqnarray}\label{TMD_crossSec_final}
d\sigma\sim F(\zeta_+)\times \bar F(\zeta_{-}),
\end{eqnarray}
where $F(\zeta_\pm)=\exp(\sigma^\pm) F^{sub}$. The function $F$ is rapidity divergences free. The procedure schematically presented here can be formulated as a kind of "rapidity renormalization", for the detailed description see \cite{Echevarria:2015usa,Luebbert:2016itl,Echevarria:2016scs,Chiu:2012ir}.

The DPS cross-section has a matrix structure (compare (\ref{TMD_crossSec_scheme}) and (\ref{DPS_cross_section_singlet})). Therefore, the rapidity factorization procedure has to be done in the matrix form. In other words DPS soft factor should be factorized onto the product of matrices with the appropriate rapidity divergences. It implies the following expression
\begin{eqnarray}\label{S=ss}
S\(\ln\(\frac{\mu^2}{\delta^+\delta^-}\)\)=s^T(\mathbf{l}^+)\cdot s(\mathbf{l}^-),
\end{eqnarray}
where $s$ is $2\times2$-matrix, and superscript $T$ denotes matrix transposition. To complete the DPS factorization formula the decomposition (\ref{S=ss}) should hold at all order of perturbation theory. However, with our current calculation we can check it only at NNLO. Using explicit expression for the matrix $S$ at NNLO (\ref{Sudud},\ref{S4},\ref{S1}) we obtain
\begin{eqnarray}\label{small_s}
s_{\uparrow\downarrow\uparrow\downarrow}(\mathbf{l}^-)=\exp\(
\begin{array}{cc}
\sigma^-_{\bf 11}
&
\sigma^-_{\bf 18}
\\
\sigma^-_{\bf 81}
&
\sigma^-_{\bf 88}
\end{array}
\),
\end{eqnarray}
where
\begin{eqnarray*}
\sigma^-_{\bf 11}&=&\sigma^-(\vec b_{14})+\sigma^-(\vec b_{23}),
\\
\sigma^-_{\bf 18}&=&\(X_1+\frac{1}{\sqrt{N_c^2-1}}\)(\sigma^-(\vec b_{12})+\sigma^-(\vec b_{34}))+\(X_2-\frac{1}{\sqrt{N_c^2-1}}\)(\sigma^-(\vec b_{13})+\sigma^-(\vec b_{24}))
\\\nn&&\qquad\qquad\qquad\qquad\qquad\qquad\qquad\qquad\qquad-(X_1+X_2)(\sigma^-(\vec b_{14})+\sigma^-(\vec b_{23})),
\\
\sigma^-_{\bf 81}&=&\(-X_1+\frac{1}{\sqrt{N_c^2-1}}\)(\sigma^-(\vec b_{12})+\sigma^-(\vec b_{34}))+\(-X_2-\frac{1}{\sqrt{N_c^2-1}}\)(\sigma^-(\vec b_{13})+\sigma^-(\vec b_{24}))
\\\nn&&\qquad\qquad\qquad\qquad\qquad\qquad\qquad\qquad\qquad+(X_1+X_2)(\sigma^-(\vec b_{14})+\sigma^-(\vec b_{23})),
\\\nn
\sigma^-_{\bf 88}&=&\frac{N_c^2-2}{N_c^2-1}(\sigma^-(\vec b_{12})+\sigma^-(\vec b_{34}))+\frac{2(\sigma^-(\vec b_{13})+\sigma^-(\vec b_{24}))}{N_c^2-1}-\frac{\sigma^-(\vec b_{14})+\sigma^-(\vec b_{23})}{N_c^2-1}.
\end{eqnarray*}
Here we use a shorthand notation
\begin{eqnarray*}
X_1&=&\frac{N_c^2}{4(N_c^2-1)^{3/2}}\(B(\vec b_{13})-B(\vec b_{14})-B(\vec b_{23})+B(\vec b_{24})\),
\\
X_2&=&\frac{N_c^2}{4(N_c^2-1)^{3/2}}\(B(\vec b_{14})-B(\vec b_{12})-B(\vec b_{34})+B(\vec b_{23})\).
\end{eqnarray*}
It is important to note, that there is a freedom in the decomposition (\ref{S=ss}) and in the definition of matrix (\ref{small_s}). We have used that freedom to make diagonal terms pure functions of $\sigma^\pm$. 

\subsection{Evolution with rapidity parameter}
\label{sec:rap_evol}

The decoupling of rapidity divergences gives rise to the dependence on rapidity parameter $\zeta$. The evolution equations with respect to the rapidity parameter are generally known as CSS equations (Collins-Soper-Sterman) \cite{Collins:1984kg}. Having at hands the soft factor one can extract the anomalous dimensions for the rapidity evolution (rAD). Let us remind this procedure in the case of TMD factorization.

The complete definition of a TMD contains several factors. The composition of these factors could be different within different formulations, compare e.g. \cite{Becher:2010tm,GarciaEchevarria:2011rb,Collins:2011zzd,Chiu:2012ir}. However, the final result for "observables" i.e. anomalous dimensions and coefficient functions, is independent on a scheme. It has been recently confirmed at NNLO by direct calculations in different schemes \cite{Echevarria:2016scs,Luebbert:2016itl,Gehrmann:2014yya}. Here we use the formulation based on $\delta$-regularization \cite{Echevarria:2015byo,Echevarria:2016scs}. Within $\delta$-regularization a TMD reads
\begin{eqnarray}\label{TMD_complete_structure}
F(\zeta_-)=Z(\zeta_-)S^{1/2}(\mathbf{l}^-)\frac{F^{unsub}(\delta^-/p^-)}{S(\mathbf{l}^-)},
\end{eqnarray}
where we drop all arguments except the argument related to rapidity parameters. The factor $Z$ is the ultraviolet renormalization constant, which is dependent on $\zeta$ via cusp-logarithms, but $\delta$-independent. The factor $S^{1/2}$ is the part of the soft factor that comes from the cross section after the procedure of rapidity divergences separation (\ref{TMD_crossSec_scheme},\ref{TMD_SF_dec},\ref{TMD_crossSec_final}). The soft factor in the denominator is the zero-bin subtractions \cite{Lee:2006nr,Manohar:2006nz}, which are equal to the soft factor in the $\delta$-regularization. Finally, $F^{unsub}$ is the TMD matrix element, that depends on $\delta$ only. In the products of these factors the regularization parameter $\delta$ cancels, and the TMD is rapidity-divergences-free. Then according to the definition of rAD
\begin{eqnarray}
\frac{d F(x,\vec b;\zeta,\mu)}{d\ln \zeta}=-\mathcal{D}(\vec b,\mu)F(x,\vec b;\zeta,\mu),
\end{eqnarray}
one has
\begin{eqnarray}
\mathcal{D}(\vec b,\mu)=-\frac{1}{2}\frac{d \ln S(\mathbf{l^-})}{d \mathbf{l}^-}-\frac{d \ln Z(\zeta)}{d \ln \zeta}.
\end{eqnarray}
The last term consists of terms that are divergent in $\epsilon$. It is needed only for the cancellation of $\epsilon$-divergences in the first term. Therefore, one can consider only the first term with all $\epsilon$-divergences dropped, and thus, can obtain rAD solely from the soft factor. Substituting the expression for the TMD soft factor (\ref{TMD_SF_dec}) we obtain
\begin{eqnarray}\label{D_TMD_from_A}
\mathcal{D}(\vec b,\mu)=-\frac{A(\vec b)}{2}\Bigg|_{f.p.},
\end{eqnarray}
where the mark $f.p.$ denotes the selection of finite part in $\epsilon$. Substituting the explicit expressions (\ref{gen_func:2V},\ref{gen_func:3V}) into (\ref{TMD_SF},\ref{TMD_SF_dec},\ref{D_TMD_from_A}) we obtain the well-known result
\begin{eqnarray}\label{def:rAD_TMD}
\mathcal{D}(\vec b,\mu)&=&2a_s(\mu)C_F\Gamma_1\mathbf{L}(\vec b\mu)
\\\nn&&
+a_s^2(\mu)C_F\(\Gamma_1\beta_0 \mathbf{L}^2(\vec b\mu)
+2 \Gamma_2\mathbf{L}(\vec b\mu)+C_A\(\frac{404}{27}-14\zeta_3\)-\frac{56}{27}N_f\)+\mathcal{O}(a_s^3),
\end{eqnarray}
where $\Gamma_1=1$ and $\Gamma_2=C_A\(\frac{67}{9}-\frac{\pi^2}{3}\)-\frac{10}{9}N_f$ are coefficients of the cusp anomalous dimension, $\beta_0=\frac{11}{3}C_A-\frac{2}{3}N_f$ is LO QCD $\beta$-function, and $\mathbf{L}(\vec b\mu)=\ln(\vec b^2 \mu^2 e^{2\gamma_E}/4)$. This coefficient can be found in many papers, see e.g. the  collection of formulae in \cite{Echevarria:2016scs}. Recently, rAD has been evaluated at NNNLO in \cite{Li:2016ctv}.

In the case of DPS the TMD scheme can be used with minimal changes. We should only take care of the matrix structure of DPD. The analog of expression (\ref{TMD_complete_structure}) in DPD case is
\begin{eqnarray}\label{DPD_complete_structure}
\bar F(\zeta_-)=Z(\zeta_-)s(\mathbf{l}^-)S^{-1}(\mathbf{l}^-)\bar F^{unsub}(\delta^-/p^-),
\end{eqnarray}
where $S^{-1}$ is the inverse matrix of soft factor responsible for zero-bin subtractions, $s$ is defined in (\ref{S=ss}), and $Z$ is the ultraviolet renormalization matrix. The order of matrices is essential and follows from the scheme of divergence recombination \cite{Echevarria:2016scs}: the rapidity divergences are canceled prior to the ultraviolet renormalization and the zero bin subtraction are the part of DPD matrix element.\footnote{The composition can be somewhat simplified within the "rapidity renormalization group" approach \cite{Chiu:2012ir}. In this case, one does not need the zero-bin subtractions, and hence rAD can be directly related to matrix $s$. The final result of both approaches is the same.} The order we set matrices in the expression (\ref{DPD_complete_structure}) corresponds to our definition of $\bar F$ as a "column". For the DPD $F$ ,which is a "row", the whole composition should be transposed.

Defining the rAD matrix in the similar way
\begin{eqnarray}
\frac{d \bar F(x_{1,2},\vec b_{1,2,3,4};\zeta,\mu)}{d\ln \zeta}=-\mathbf{D}(\vec b_{1,2,3,4},\mu)\bar F(x_{1,2},\vec b_{1,2,3,4};\zeta,\mu),
\end{eqnarray}
we obtain
\begin{eqnarray}\label{DD_as_derivative}
\mathbf{D}(\vec b_{1,2,3,4},\mu)=-\frac{d Z(\zeta)}{d\ln \zeta}Z^{-1}+Z(\zeta)\frac{d (s^{T}(\mathbf{l^-}))^{-1}}{d\mathbf{l}^-}s^{T}(\mathbf{l^-})Z(\zeta)
=\frac{d (s^{T}(\mathbf{l^-}))^{-1}}{d\mathbf{l}^-}s^{T}(\mathbf{l^-})\Bigg|_{f.p.}~.
\end{eqnarray}
To obtain the last equality we have commuted the matrix $Z$ to the right and dropped singular in $\epsilon$ terms.

It appears, that the expression for the rAD matrix can be found without referring to explicit expressions for loop integrals. Substituting the matrix $\sigma^-$ in the form (\ref{small_s}) into the equation (\ref{DD_as_derivative}), we obtain the expression which consists entirely of function $A$ in the form (\ref{D_TMD_from_A}). Therefore, the rADs for DPDs can be expresses via the TMD rAD (\ref{def:rAD_TMD})
\begin{eqnarray}\label{rAD1}
&&\mathbf{D}_{q\bar q}(\vec b_{1,2,3,4},\mu)=\mathbf{D}_{\bar q q}(\vec b_{1,2,3,4},\mu)
\\\nn&&=\(
\begin{array}{cc}
\mathcal{D}(\vec b_{14})+\mathcal{D}(\vec b_{23}) & \frac{\mathcal{D}(\vec b_{12})-\mathcal{D}(\vec b_{13})-\mathcal{D}(\vec b_{24})+\mathcal{D}(\vec b_{34})}{\sqrt{N_c^2-1}}
\\
\frac{\mathcal{D}(\vec b_{12})-\mathcal{D}(\vec b_{13})-\mathcal{D}(\vec b_{24})+\mathcal{D}(\vec b_{34})}{\sqrt{N_c^2-1}}
&
\frac{N^2_c-2}{N^2_c-1}\(\mathcal{D}(\vec b_{12})+\mathcal{D}(\vec b_{34})\)+\frac{2(\mathcal{D}(\vec b_{13})+\mathcal{D}(\vec b_{24}))-\mathcal{D}(\vec b_{14})-\mathcal{D}(\vec b_{23})}{N_c^2-1}
\end{array}
\),
\end{eqnarray}
where we drop the argument $\mu$ from the function $\mathcal{D}$ for brevity. The rADs for $F_{qq}$, $F_{\bar q \bar q}$, as well as, for $\mathcal{I}_{q\bar q}$ and $\mathcal{I}_{\bar q q}$ can be obtained by permutation of vectors $\vec b$ according to (\ref{sf_relations1}-\ref{sf_relations4},\ref{sf_relations5})
\begin{eqnarray}\label{rAD2}
\mathbf{D}_{q q}(\vec b_1,\vec b_2,\vec b_3,\vec b_4)&=&\mathbf{D}_{\bar q \bar q}(\vec b_1,\vec b_2,\vec b_3,\vec b_4)=
\mathbf{D}_{q \bar q}(\vec b_1,\vec b_3,\vec b_2,\vec b_4),
\\
\mathbf{D}^{\mathcal{I}}_{q \bar q}(\vec b_1,\vec b_2,\vec b_3,\vec b_4)&=&\mathbf{D}^{\mathcal{I}}_{\bar q q}(\vec b_1,\vec b_2,\vec b_3,\vec b_4)=
\mathbf{D}_{q \bar q}(\vec b_1,\vec b_2,\vec b_4,\vec b_3).
\end{eqnarray}
There are some elementary checks of these expressions. First of all, these expressions do not contain $\delta$-dependence, it cancel in the product of matrices (\ref{DD_as_derivative}). Second, the matrices $\mathbf{D}$ are symmetric matrices (although the matrix $s$ is not), it implies that DPDs $F$ and DPDs $\bar F$ evolve by the same equations, which in turn is the requirement of Lorentz invariance.

The final result for the rADs at NNLO is the same (in pattern) as at the NLO \cite{Diehl:2011yj}. In this way, the only difference from the result for NLO rapidity evolution given in \cite{Diehl:2011yj}, is that TMD rAD $\mathcal{D}$ should be taken at NNLO. This conclusion is very natural since any three Wilson line correlations disappear. However, this pattern does not hold at NNNLO where correlations of four Wilson lines appear. These correlators does not cancel in the sum of diagrams and give rise to a new ``quadropole'' contribution. \footnote{In the recent paper \cite{Vladimirov:2016dll} the NNNLO expression for the rapidity anomalous dimension $\mathbf{D}$ has been obtained, using the conformal correspondence between rAD and the soft anomalous dimension. The NNLO expression for rAD indeed contains the quadrapole contributions, together with the TMD rAD.}

\section{Expressions for integrated kinematics}
\label{sec:integrated}

In this section we collect the results for the integrated double-parton scattering. All presented results are obtained by from the expressions of the previous sections.

The integrated double-Drell-Yan cross-section is obtained from the integrated one (\ref{DPS_cross_section}) by the integration over momenta $\vec q_{1,2}$. In the regime $\Lambda_{QCD}^2\ll s$ the factorized expression is
\begin{eqnarray}\label{DPS_cross_section_int}
\frac{d\sigma}{d X}\Big|_{\text{DPS}}&=&d\tilde \sigma_{\{ij\}}\int d^2\vec y \Bigg[
f^{b_1b_2a_3a_4}_{qq,\{ij\}}\bar f^{a_1a_2b_3b_4}_{\bar q\bar q,\{ij\}}S^{\{ab\}}_{\uparrow\uparrow\downarrow\downarrow}(\vec y)+
f^{a_1a_2b_3b_4}_{\bar q\bar q,\{ij\}}\bar f^{b_1b_2a_3a_4}_{qq,\{ij\}}S^{\{ab\}}_{\downarrow\downarrow\uparrow\uparrow}(\vec y)
\\\nn&&\qquad\qquad\qquad
f^{b_1a_2b_3a_4}_{q\bar q,\{ij\}}\bar f^{a_1b_2a_3b_4}_{\bar q q,\{ij\}}S^{\{ab\}}_{\uparrow\downarrow\uparrow\downarrow}(\vec y)+
f^{a_1b_2a_3b_4}_{\bar q q,\{ij\}}\bar f^{b_1a_2b_3a_4}_{q\bar q,\{ij\}}S^{\{ab\}}_{\downarrow\uparrow\downarrow\uparrow}(\vec y)
\\\nn&&\qquad\qquad\qquad
I^{b_1a_2a_3b_4}_{\bar qq,\{ij\}}\bar I^{a_1b_2b_3a_4}_{q\bar q,\{ij\}}S^{\{ab\}}_{\uparrow\downarrow\downarrow\uparrow}(\vec y)+
I^{a_1b_2b_3a_4}_{q\bar q,\{ij\}}\bar I^{b_1a_2a_3b_4}_{\bar qq,\{ij\}}S^{\{ab\}}_{\downarrow\uparrow\uparrow\downarrow}(\vec y)
\Bigg],
\end{eqnarray}
where phase-space element is $d X=dx_1d\bar x_1~dx_2d\bar x_2=s^{-2}dq^2_1dY_1~dq^2_2dY_2$. The integrated DPDs are related to unintegrated as
\begin{eqnarray}\label{DPD_to_int}
f(x_{1,2},\vec y)=F(x_{1,2},\vec y,\vec 0,\vec 0,\vec y),\qquad 
I(x_{1,2},\vec y)=\mathcal{I}(x_{1,2},\vec y,\vec 0,\vec 0,\vec y),
\end{eqnarray}
where we drop all indices for brevity, and functions $F(x_{1,2},\vec b_1,\vec b_2,\vec b_3,\vec b_4)$ are defined in equations (\ref{DPD_TMD_def1},\ref{DPD_TMD_def2},\ref{DPD_TMD_def3}). Note, that integrated DPDs $F$ and $I$ have essentially different structure. The DPDs $F$ have quark-antiquark pair created at the same traverse coordinate, which effectively represents a gluon creation. The DPDs $I$ have a quark-quark pair created at the transverse distance. It is reflected in the very different structure of soft factor (\ref{SF_integrated1}-\ref{SF_integrated2}). 

The integrated soft factors are
\begin{eqnarray}\label{SF_to_int}
S(\vec y)=S(\vec y,\vec 0,\vec 0,\vec y),
\end{eqnarray}
where we drop all indices for brevity, and functions $S(\vec b_1,\vec b_2,\vec b_3,\vec b_4)$ are defined in equation (\ref{SF_gen_def}). Important to note that the relations (\ref{DPD_to_int},\ref{SF_to_int}) are formal relations. The singularity structures on the left- and right-hand sides of equations (\ref{DPD_to_int},\ref{SF_to_int}) are different. To properly match the singularities the limit $\vec b\to 0$ should be performed before the removal of regulator.

Contrary to the usual intuition for single Drell-Yan processes, the integrated soft factor matrices are not unity matrices. Correspondingly the integrated DPS contains the rapidity divergences, which are associated with vector $\vec y$ (\ref{DPS_cross_section_int}). The transverse vector $\vec y$ is not observable, and intrinsic for the factorization formula (\ref{DPS_cross_section_int}). 

The soft factors in the integrated case can be obtained from the unintegrated ones by the relation (\ref{SF_to_int}). Using that $\sigma^\pm(\vec 0)=0$ we find
\begin{eqnarray}\label{SF_integrated1}
S_{\uparrow\downarrow\uparrow\downarrow}(\vec y)&=&
S_{\downarrow\uparrow\downarrow\uparrow}(\vec y)=S_{\uparrow\uparrow\downarrow\downarrow}(\vec y)=S_{\downarrow\downarrow\uparrow\uparrow}(\vec y)=\exp \(
\begin{array}{cc}
0&0
\\
0 &\frac{C_A}{C_F}\sigma(\vec y)
\end{array}\),
\\\label{SF_integrated2}
S_{\uparrow\downarrow\downarrow\uparrow}(\vec y)&=&S_{\downarrow\uparrow\uparrow\downarrow}(\vec y)=\exp \(
\begin{array}{cc}
2\sigma(\vec y)& \frac{2 \sigma(\vec y)}{\sqrt{N_c^2-1}}
\\
\frac{2 \sigma(\vec y)}{\sqrt{N_c^2-1}} &2\frac{N_c^2-3}{N_c^2-1}\sigma(\vec y)
\end{array}\).
\end{eqnarray}
Naturally integrated soft factors are expressed via the TMD soft factor. The matrices $s$ are obtained by replacing $\sigma(\vec y)\to \sigma^-(\vec y)$, i.e.
\begin{eqnarray}
s(\mathbf{l}^-)=S[\sigma^-].
\end{eqnarray}

The relations (\ref{SF_integrated1}-\ref{SF_integrated2}) must hold to high orders of perturbation theory and, in fact, represents the so-called Casimir scaling for a Wilson loop (i.e. expressions for Wilson loop of different representations differ only by the overall Casimir eigenvalue). Indeed, gluing together $\Lambda$ in $\bf 8\bf 8$ element we obtain a single $\Lambda$ in the adjoint representation, or TMD soft factor in the adjoint representation. However, it is possible that at four- or higher loop order this relation would be violated by the qubic Casimirs terms.

The anomalous dimension matrices for the integrated case are obtained by the same procedure as for integrated. Considering (\ref{DD_as_derivative}) with soft factors (\ref{SF_integrated1}-\ref{SF_integrated2}) we obtain
\begin{eqnarray}\label{rAD3}
\mathbf{D}_{q q}(\vec y)&=&\mathbf{D}_{\bar q \bar q}(\vec y)=\mathbf{D}_{\bar q q}(\vec y)=\mathbf{D}_{q \bar q}(\vec y)=
\(\begin{array}{cc}
0 & 0
\\ 0 & \frac{C_A}{C_F} \mathcal{D}(\vec y)
\end{array}\),
\\\label{rAD4}
\mathbf{D}^{\mathcal{I}}_{q \bar q}(\vec y)&=&\mathbf{D}^{\mathcal{I}}_{\bar q q}(\vec y)=
\(\begin{array}{cc}
2\mathcal{D}(\vec y) & \frac{2\mathcal{D}(\vec y)}{\sqrt{N_c^2-1}}
\\ \frac{2 \mathcal{D}(\vec y)}{\sqrt{N_c^2-1}} & 2\frac{N_c^2-3}{N_c^2-1} \mathcal{D}(\vec y)
\end{array}\).
\end{eqnarray}
One can see that the same expression can be obtained from (\ref{rAD1}-\ref{rAD2}) if we assume $\mathcal{D}(\vec 0)=0$.

\section{Conclusion}

We have considered the soft factor for the leading order factorization formula of the double parton scattering (DPS) in  the perturbative regime (such that $\ln( \vec b^2 Q^2)$ is not large, where $\vec b$ is any transverse separation within the soft factor).  We have shown that at NNLO the soft factors  are expressed entirely via the soft factor for transverse momentum dependent (TMD) factorization. This simplification happens due to the exact cancellation of the non-trivial part of three-Wilson lines interaction, while the trivial part of the three-Wilson line interaction can be presented via the pairwise interaction. Therefore, the DPS soft factor, that is generically a function of four transverse vectors, at NNLO reduces to products of simple functions of a single variable.

Using the fact that the logarithm of TMD soft factor is a linear function of rapidity divergence the DPS soft factor matrix can be split into a product of matrices that contain only the rapidity divergences in appropriate sectors (\ref{S=ss}). Adjusting the matrices to the unsubtracted double parton distributions (DPDs)(in other words, singular DPDs) we define DPDs that are free of rapidity divergences. It validates the DPD factorization theorem at NNLO. Using explicit expression for the soft factor matrices we extract the expressions for the rapidity anomalous dimension (rAD) matrices. At NNLO, the rAD matrices are expressed via rAD for TMD distributions (\ref{rAD1},\ref{rAD4}). It appears that the expression for NNLO rAD follows the pattern of NLO expression with the only substitution of the two-loop anomalous dimension. Such simple pattern of DPS soft factor does not hold at higher orders. Starting from the NNNLO new functions appear. These functions represent the simultaneous interaction of four Wilson lines (quadrupole terms). 

To express the DPS soft factor via TMD soft factors we used the generating function for web diagrams \cite{Vladimirov:2015fea,Vladimirov:2014wga}. This is the first application of the method to a previously unknown object. This calculation shows high efficiency of the generating function approach. In fact, to obtain many results one does not need any explicit expression for diagrams but only the symmetry properties of Wilson lines which became transparent in the generating function approach. Using explicit expression for the generating function in the $\delta$-regularization (presented in appendix \ref{app:calc}) we confirm the results of \cite{Echevarria:2015byo}. As a side result, we present the generating function for web diagrams at NNLO for Wilson lines with two light-like directions, which can be used to obtain any matrix element of Wilson lines with similar geometry, e.g. a soft factor for multi-parton scattering.

The presented calculation has been done in the Drell-Yan kinematics. In this case, all Wilson lines of the soft factor can be collected in the single T-ordered product. This trick reduces the number of diagrams and simplifies the calculation. In the case of Wilson lines pointing in different time directions (e.g. double semi-inclusive deep-inelastic scattering (SIDIS) kinematics), one should additionally consider the diagrams with real-gluon-exchange. However, we do not expect any special contribution from these diagrams at NNLO. The point is that the TMD soft factor is the same in both kinematics. This property is known as the universality of the TMD soft factor, see \cite{Collins:2004nx,Echevarria:2015byo}. Therefore, we conclude that the DPD soft factor is also universal at least at NNLO.

We stress that relations between DPD soft factors and TMD soft factors are based on the geometry of the construction and color algebra only. Therefore, the results of the article are independent of the regularization scheme (although we imply some basic "good" properties of a regularization, such as non-violation of the exponentiation theorem).

\acknowledgments The author gratefully acknowledges V.Braun and I.Scimemi for numerous stimulating discussions, and M.Diehl for useful comments. Also, the author acknowledges Erwin Schrödinger International Institute for Mathematics and Physics (ESI) for the support during the programme ``Challenges and Concepts for Field Theory
and Applications in the Era of LHC Run-2'', which stimulates this research.

\appendix

\section{Relation between normalizations}
\label{app:normalizations}

There are several sets of notations used for DPDs. We write the explicit relation of DPDs considered in this work to those introduced in \cite{Manohar:2012jr} and in \cite{Diehl:2011yj}. Since we did not consider the Lorentz structure we leave it hidden and discuss only the color decomposition and the order of transverse arguments.

In ref.\cite{Manohar:2012jr} only the integrated kinematics is considered. The color decomposition has different normalization (see equation (56)). We found
\begin{eqnarray}
F^1(\vec z_\perp)=F^{\bm 1}(\vec z_\perp,\vec 0,\vec 0,\vec z_\perp),\qquad F^T(\vec z_\perp)=\frac{\sqrt{N_c^2-1}}{2 N_c}F^{\bm 8}(\vec z_\perp,\vec 0,\vec 0,\vec z_\perp),
\end{eqnarray}
where $F$ is DPD of any quark configuration, including $\mathcal{I}$. Comparing equations (44-49) we found the following relations 
\begin{eqnarray}\nn
S^{11}(\vec z_\perp)&=&S^{[2]}(\vec z_\perp,\vec z_\perp,\vec 0,\vec 0)=1,
\\\nn
S^{TT}(\vec z_\perp)&=&\frac{N_c^2}{N_c^2-1}(S^{[1]}(\vec z_\perp,\vec z_\perp,\vec 0,\vec 0)-S^{[2]}(\vec z_\perp,\vec z_\perp,\vec 0,\vec 0)),
\\
S^{11}_I(\vec z_\perp)&=&S^{[2]}(\vec z_\perp,\vec z_\perp,\vec 0,\vec 0),
\\\nn
S^{T1}_I(\vec z_\perp)&=&\frac{-2S^{[2]}(\vec z_\perp,\vec z_\perp,\vec 0,\vec 0)}{N_c^2-1}+\frac{2N_cS^{[4]}(\vec z_\perp,\vec 0,\vec z_\perp,\vec 0)}{N_c^2-1},
\\
S^{TT}_I(\vec z_\perp)&=&\frac{N_c^2+1}{N_c^2-1}S^{[2]}(\vec z_\perp,\vec z_\perp,\vec 0,\vec 0)-\frac{2N_cS^{[4]}(\vec z_\perp,\vec 0,\vec z_\perp,\vec 0)}{N_c^2-1}.
\end{eqnarray}

The definitions of DPDs used in \cite{Diehl:2011yj} coincide with our definition (see equation (2.103) and (2.104)) with the following adjustment of transverse arguments
\begin{eqnarray}\nn
\,^{1,8}F_{q q}(x_{1,2},\vec y,\vec z_{1,2})=F_{q q}^{\bm 1,\bm 8}(x_{1,2},\vec y+\frac{\vec z_2}{2},\frac{\vec z_2}{2},-\frac{\vec z_2}{2},\vec y-\frac{z_1}{2}),
\\
\,^{1,8}F_{q \bar q}(x_{1,2},\vec y,\vec z_{1,2})=F_{q \bar q}^{\bm 1,\bm 8}(x_{1,2},\vec y+\frac{\vec z_2}{2},\frac{\vec z_2}{2},-\frac{\vec z_2}{2},\vec y-\frac{z_1}{2}),
\\\nn
\,^{1,8}\mathcal{I}_{q \bar q}(x_{1,2},\vec y,\vec z_{1,2})=\mathcal{I}_{q \bar q}^{\bm 1,\bm 8}(x_{1,2},\vec y+\frac{\vec z_2}{2},\frac{\vec z_2}{2},-\frac{\vec z_2}{2},\vec y-\frac{z_1}{2}).
\end{eqnarray}
Correspondence between soft factors is following 
\begin{eqnarray}
S_{q\bar q}(\vec y,\vec z_{1,2})&=&S_{\uparrow\downarrow\uparrow\downarrow}\(\vec y+\frac{\vec z_2}{2},\frac{\vec z_2}{2},-\frac{\vec z_2}{2},\vec y-\frac{z_1}{2}\),
\\
S_{q q}(\vec y,\vec z_{1,2})&=&S_{\uparrow\uparrow\downarrow\downarrow}\(\vec y+\frac{\vec z_2}{2},\frac{\vec z_2}{2},-\frac{\vec z_2}{2},\vec y-\frac{z_1}{2}\),
\\
S_{I}(\vec y,\vec z_{1,2})&=&S_{\uparrow\downarrow\downarrow\uparrow}\(\vec y+\frac{\vec z_2}{2},\frac{\vec z_2}{2},-\frac{\vec z_2}{2},\vec y-\frac{z_1}{2}\).
\end{eqnarray}

\section{Details of NNLO calculation}
\label{app:calc}

In this appendix we present the details of generating function evaluation. Also we demonstrate explicitly some statements discussed in section \ref{sec:2loop}. The numeration of diagrams follow the fig.\ref{fig:2loop}.

\subsection{Diagram $\langle 2 2 \rangle$}

The diagram $\langle 2 2 \rangle$ has the symmetry factor $1/2$. The Feynman rules for $V_2$ vertex are given in (\ref{V2_FeynRules}). The explicit expression for diagram is
\begin{eqnarray}
\langle V_2 V_2 \rangle&=&\frac{1}{2}v_{ij}^2\theta_i^A\theta_j^{A'}\frac{g^4}{4}if^{ABC}if^{A'BC}\int \frac{d^dkd^dl}{(2\pi)^{2d}}
\frac{-e^{i(\vec k+\vec l)\vec b_{ji}}}{(k^2+i0)(l^2+i0)}
\\&&
\frac{1}{(-k_i-l_i+2i\delta_i)(k_j+l_j+2i\delta_j)}\(\frac{1}{-k_i+i\delta_i}-\frac{1}{-l_i+i\delta_i}\)
\(\frac{1}{k_j+i\delta_j}-\frac{1}{l_j+i\delta_j}\) \nn.
\end{eqnarray}
Here and further we use the shorthand notation for scalar products $k_i=(k \cdot v_i)$.  Rewriting in the terms of base integrals we obtain
\begin{eqnarray}\label{app:eqn_22}
\langle V_2 V_2 \rangle=\frac{v_{ij}C_A}{2}g^4 \theta_i^A\theta_j^A\(\frac{(K^{(0)}(\vec b_{ij}))^2}{4}-I''_A(\vec b_{ij})\),
\end{eqnarray}
where integral $I''_{A}$ are given in (\ref{app:Int_IA}). One can check that at $\vec b_{ij}=\vec 0$ the expression (\ref{app:eqn_22}) turns to zero. It is the consequence of the hidden symmetry of the generating function on light-like Wilson lines \cite{Vladimirov:2015fea}.

\subsection{Diagram $\langle 2 1 \rangle$}

The diagram $\langle 2 1 \rangle$ has symmetry factor $1/2$. The explicit expression for diagrams is
\begin{eqnarray*}
\langle V_2 V_1 \rangle&=&\frac{g^4}{4}v_{ij}\theta_i^A\theta_j^{A'}if^{ABC}if^{CBA'}~\int \frac{d^dkd^dl}{(2\pi)^{2d}}
\\&&
\frac{e^{i\vec k\vec b_{ij}}(2l_i+k_i)}{(k_j+i\delta_j)(-k_i+2i\delta_j)(k^2+i0)(l^2+i0)((k+l)^2+i0)}\(\frac{1}{l_i+i\delta_i}-\frac{1}{-k_i-l_i+i\delta_i}\).
\end{eqnarray*}
The second term in brackets is equal the first one under the change of variables $l\to -l-k$. The configuration of eikonal propagators can be rewritten as (we set $i=+$ and $j=-$ for transparency)
\begin{eqnarray}
\frac{2l_i+k_i}{(k_j+i\delta_j)(-k_i+2i\delta_j)(l_i+i\delta_i)}=\frac{1}{k^-+i\delta^-}\(\frac{2}{k^+-2i\delta^+}+\frac{1}{l^++i\delta^+}\).
\end{eqnarray}
In terms of base integrals we obtain
\begin{eqnarray}
\langle V_2 V_1 \rangle=-\frac{v_{ij}C_A}{2}g^4 \theta_i^A\theta_j^A\(2 I_{C1}(\vec b_{ij})+I_{C2}(\vec b_{ij})\).
\end{eqnarray}

The diagram $\langle 21 \rangle$ has an ultravioletly divergent subgraph to be renormalized. The renormalization of this subgraph corresponds to the expansion of the renormalization factor $Z_gZ^{1/2}$ of the one-loop diagram. The counter term is
\begin{eqnarray}
\langle 2 1\rangle_{CT}=-\theta^A_i\theta^A_jv_{ij}a_s g^2
\frac{2C_A}{\epsilon}K^{(0)}(\vec b_{ij}).
\end{eqnarray}

\subsection{Diagram $\langle 1 1 \rangle$}

The evaluation of diagram $\langle 1 1 \rangle$ is straightforward, and in all aspects repeats the one done in \cite{Echevarria:2015byo}. Therefore, we present only the final expression
\begin{eqnarray}
\langle V_1 V_1 \rangle
 &=& -\frac{a_s^2\theta^A_i\theta^A_jv_{ij}}{2}\frac{4\Gamma(2-\epsilon)\Gamma(-\epsilon)\Gamma(1+\epsilon)}{\Gamma(4-2\epsilon)}
\(C_A(5-3\epsilon)-2N_f(1-\epsilon)\) K^{(\epsilon)}(\vec b_{ij})
\\\nn&&-\theta^A_i\theta^A_jv_{ij}a_sg^2
\frac{1}{\epsilon}\(\frac{5}{3}C_A-\frac{2}{3}N_f\)K^{(0)}(\vec b_{ij}),
\end{eqnarray}
where the last line is the counter term.

\subsection{Diagram $\langle 2 1 1 \rangle$}

The expression for diagram is
\begin{eqnarray*}
\langle V_2V_1V_1\rangle&=&\frac{g^4}{2}\theta^A_i\theta^B_j\theta^C_k v_{ij}v_{ik} if^{ABC}\int \frac{d^dkd^dl}{(2\pi)^{2d}}
\frac{-e^{i\vec k\vec b_{ji}}e^{i\vec l\vec b_{ki}}}{(k^2+i0)(l^2+i0)}
\\&&
\frac{1}{(k_j+i\delta_j)(l_k+i\delta_k)(-k_i-l_i+2i\delta_i)}\(\frac{1}{-l_i+i\delta_i}-\frac{1}{-k_i+i\delta_i}\).
\end{eqnarray*}
The second term in the brackets is related to the first one by the replacement $k\leftrightarrow l$. Therefore, we need to consider only the following integral
\begin{eqnarray*}
J(\vec b_1,\vec b_2)&=&
\int \frac{d^dkd^dl}{(2\pi)^{2d}}
\frac{-e^{i\vec k\vec b_1}e^{i\vec l\vec b_2}}{k^2l^2(k^++i\delta^+)(l^++i\delta^+)(k^-+l^--2i\delta^-)(l^--i\delta^-)},
\end{eqnarray*}
here $k^2=k^2+i0$. In this notation the diagram reads
\begin{eqnarray}
\langle V_2V_1V_1\rangle&=&\frac{g^4}{2}\theta^A_i\theta^B_j\theta^C_k v_{ij}v_{ik} if^{ABC}\(J(\vec b_{ij},\vec b_{ik})-J(\vec b_{ik},\vec b_{ij})\),
\end{eqnarray}
where expression for integral $J$ is given in (\ref{app:Int_J}). Recalling that the generating function contains the sum of the diagrams with operator $V_2$ on different lines we obtain the contribution to $W_{ijk}$ in the form
\begin{eqnarray}
W^{\langle 211\rangle}_{ijk}&=&\frac{a_s^2}{2}\Bigg[v_{ij}v_{ik}(J(\vec b_{ij},\vec b_{ik})-J(\vec b_{ik},\vec b_{ij}))
+v_{ij}v_{jk}(J(\vec b_{jk},\vec b_{ij})-J(\vec b_{ij},\vec b_{jk}))
\\ \nn &&+v_{jk}v_{ik}(J(\vec b_{ik},\vec b_{jk})-J(\vec b_{jk},\vec b_{ik}))
\Bigg].
\end{eqnarray}
This representation is totally antisymmetric under the permutation of indices $ijk$. Taking into account the antisymmetry of prefactor  $\theta^A_i\theta^B_j\theta^C_k if^{ABC}$ we write it in the simple form, which is used in (\ref{def:Wijk})
\begin{eqnarray}
W^{\langle 211\rangle}_{ijk}=3 a_s^2 v_{ij}v_{ik}J(\vec b_{ij},\vec b_{ik}).
\end{eqnarray}
At $\vec b_{ij}=\vec b_{ik}=\vec 0$ the integral $J$ turns to zero in accordance to symmetries of the generating function.

\subsection{Diagram $\langle 1 1 1 \rangle$}

The expression for the diagram $\langle 1 1 1 \rangle$ is
\begin{eqnarray*}
\langle V_1V_1V_1\rangle&=&g^4\theta^A_i\theta^B_j\theta^C_k if^{ABC}
\int \frac{d^dkd^dl}{(2\pi)^{2d}}e^{i\vec k\vec b_{ij}}e^{i\vec l\vec b_{ik}}
\\&&
\frac{(2k_k+l_k)v_{ij}+(l_i-k_i)v_{jk}-(2l_j+k_j)v_{ik}}{(-k_i-l_i+i\delta_i)(k_j+i\delta_j)(l_k+i\delta_k)(k^2+i0)(l^2+i0)((k+l)^2+i0)}.
\end{eqnarray*}
Contrary to the previous diagrams within diagram $\langle 1 1 1 \rangle$ there are multiple possibilities to associate vectors $v_i$, $v_j$, and $v_k$ with $n$ and $\bar n$. For example, the term with $v_{ij}$ requires only $v_i\neq v_j$, while does not fix $v_k$. The convenient way to consider this diagram is to write explicitly both possibilities for vector $v_k$ (i.e. $v_k=v_i$ and $v_k=v_j$) as  $v_{ij}(v_{ik}+v_{jk})$. Then the numerator can be simplified
\begin{eqnarray}\nn
(2k_k+l_k)v_{ij}+(l_i-k_i)v_{jk}-(2l_j+k_j)v_{ik}=\!\!(-2k_i-l_i)v_{ik}v_{jk}+(k_j-l_j)v_{ij}v_{ik}+(k_i+2l_i)v_{ij}v_{jk}.
\end{eqnarray}
Using this trick and change of variables (alike $k\to -k-l$) in some terms we obtain the following expressions
\begin{eqnarray}\label{app:label1}
\langle V_1V_1V_1\rangle&=&g^4\theta^A_i\theta^B_j\theta^C_k if^{ABC}\Big[
R(\vec b_{ij},\vec b_{ik})(v_{ij}v_{jk}-v_{ik}v_{jk})
\\\nn &&+R(\vec b_{ij},\vec b_{kj})(v_{ik}v_{jk}-v_{ij}v_{ik})
+R(\vec b_{ik},\vec b_{jk})(v_{ij}v_{ik}-v_{ij}v_{jk})\Big],
\end{eqnarray}
where
\begin{eqnarray*}
R(\vec b_1,\vec b_2)&=&\int \frac{d^dkd^dl}{(2\pi)^{2d}}\frac{-e^{i\vec k\vec b_1}e^{i\vec l\vec b_2}}{(k^++i\delta^+)(l^-+i\delta^-)k^2l^2(k+l)^2}.
\end{eqnarray*}
To obtain expression (\ref{app:label1}) we have also used the symmetries of the integral
\begin{eqnarray}\label{app:R_sym}
R(\vec b_1,\vec b_2)=R(\vec b_2,\vec b_1),\qquad
R(-\vec b_1,-\vec b_2)=R(\vec b_1,\vec b_2),
\end{eqnarray}
The expression (\ref{app:label1}) is totally anti-symmetric under the permutation of indices $ijk$. Using it we rewrite the diagram $\langle 1 1 1 \rangle$ in the form
\begin{eqnarray}
W^{\langle 111\rangle}_{ijk}=6 a_s^2 v_{ij}v_{ik}R(\vec b_{ik},\vec b_{jk}).
\end{eqnarray}
The expression for integral $R$ is given in (\ref{app:Int_R2}).

\section{Expression for loop integrals}
\label{app:loop_integrals}

It appears that one needs less number of base loop-integrals in comparison to the calculation presented in \cite{Echevarria:2015byo}, but the integrals are of more general structure. In fact, the most part of integrals evaluated in \cite{Echevarria:2015byo} are particular cases of a few general integrals.

The loop-integrals presented in the following paragraphs are all taken by the following strategy:
\begin{itemize}
\item The integral over convenient light-cone components, say $k^+$ and $l^+$ are taken by residues closing the integration contour in lower- or upper-half of the complex plane. It also restricts the integrations over opposite light-cone components.
\item The obtained propagators are expanded in Mellin-Barnes (MB) representation such that transverse components are separated from the light-cone ones.
\item The integration over the transverse components is reduced to the scalar massless two-loop integrals in $2-2\epsilon$  dimensions. Except the integral $R$ these scalar integrals are product of $\Gamma$-functions only.  
\item The integrals over the rest light-cone components are table integrals of hypergeometric type. At this point of evaluation the expression for integral has form of single- or double-MB integral over product of $\Gamma$-function and possibly additional hypergeometric function. Note, that along all evaluation the regularization parameter $\epsilon >0$.
\item The MB integration is taken by closing the contour. This integration is significantly simplified by the fact that we need only the leading small-$\delta$ asymptotic. It allows to consider only the poles in the vicinity of zero for many series of poles. The droped terms correspond to the power suppressed terms in the small-$\delta$ expansion.
\end{itemize}

\subsection{Two point integrals}

The two-point integrals appear in the diagrams that connect only two Wilson lines and contribute into $W_{ij}$. All these integrals are calculated in \cite{Echevarria:2015byo}, and can be presented in the closed form in terms of hypergeometric function and its derivatives.

Generic one-loop integral is
\begin{eqnarray}\label{app:int_K}
K^{(a)}[\vec b]&=&(4\pi)^d\int \frac{d^d k}{(2\pi)^d}\frac{-ie^{i\vec k\vec b}}{(k^++i\delta^+)(k^--i\delta^-)(-k^2-i0)^{1+a}},
\\\nn &=&
\frac{2}{(4\pi)^{d/2}}\(\pmb \delta^{-a-\epsilon}\frac{\Gamma^2(a+\epsilon)\Gamma(1-a-\epsilon)}{\Gamma(1+a)}
-\pmb B^{a+\epsilon}\frac{\Gamma(-a-\epsilon)}{\Gamma(1+a)}\(\mathbf{L}_\delta -\mathbb{S}_{a+\epsilon}^1\)\),
\end{eqnarray}
where
\begin{eqnarray}
\mathbb{S}^1_x&=&S_1(1-x)=\psi(-x)+\gamma_E,\\\nn
\mathbb{S}^k_x&=&S_k(1-x)=\frac{(-1)^{k-1}}{(k-1)!}\psi^{(k-1)}(-x)+\zeta_k,
\end{eqnarray}
\begin{eqnarray}\nn
\pmb B_{ij}=\frac{\vec b_{ij}^2}{4} >0,~~~\pmb \delta=2\delta^+\delta^-,
\end{eqnarray}
\begin{eqnarray}\nn
\mathbf{L}_\delta&=&\ln \(\pmb B \pmb \delta e^{2\gamma_E}\)=\ln \(\frac{\vec b^2}{4}\frac{2\delta^+\delta^-}{e^{-2\gamma_E}}\).
\end{eqnarray}

The two-loop integrals are
\begin{eqnarray}\label{app:Int_IA}
I''_A(\vec b)&=&\int \frac{d^d k d^d l}{(2\pi)^{2d}}\frac{(2\pi)^2\delta_+(k^2)\delta_+(l^2)e^{i(\vec k+\vec l)\cdot \vec b}}{(l^++i\delta^+)(k^++l^++2i\delta^+)(k^-+i\delta^-)(k^-+l^-+2i\delta^-)}
\\\nn  &=&\frac{\pmb \delta^{-2\epsilon}}{(4\pi)^d}\Gamma^4(\epsilon)\Gamma^2(1-\epsilon)-4\frac{\pmb B^{\epsilon}\pmb \delta^{-\epsilon}}{(4\pi)^d}F_\epsilon-8\frac{\pmb B^{2\epsilon}}{(4\pi)^d}\(\Gamma(-2\epsilon)\Gamma(-\epsilon)\Gamma(\epsilon)\psi\(\frac{1-\epsilon}{2}\)+Q(\epsilon)\)
\\\nn &&+\frac{\pmb B^{2\epsilon}}{(4\pi)^d}\Gamma^2(-\epsilon)\Big[2\(\mathbf{L}_\delta+\mathbb{S}^1_{2\epsilon}-2\mathbb{S}^1_{\epsilon}\)^2-8\(\ln 2+\gamma_E\)\(\mathbb{S}^1_{\epsilon}-\mathbb{S}^1_{2\epsilon}\)+6 \mathbb{S}^2_{2\epsilon}-4 \mathbb{S}^2_{\epsilon}-4\zeta_2\Big],
\end{eqnarray}
\begin{eqnarray}\label{app:Int_IC1}
I_{C1}(\vec b)&=&\int \frac{d^dkd^dl}{(2\pi)^{2d}}\frac{e^{ikb}}{k^2l^2(k+l)^2(k^-+i\delta^-)(k^+-2i\delta^+)}=
\frac{\Gamma(\epsilon)\Gamma^2(1-\epsilon)}{\Gamma(2-2\epsilon)}K^{(\epsilon)}(\vec b)\Big|_{\pmb \delta \to 2 \pmb \delta},
\\\label{app:Int_IC2}
I_{C2}(\vec b)&=&\int \frac{d^dkd^dl}{(2\pi)^{2d}}\frac{e^{ikb}}{k^2l^2(k+l)^2(k^-+i\delta^-)(l^++i\delta^+)}
=\frac{2}{(4\pi)^{d}}\Bigg[\pmb B^\epsilon \pmb \delta^{-\epsilon}\Gamma^2(\epsilon)\Gamma^2(-\epsilon)
\\&&\nn+\pmb \delta^{-2\epsilon}\Gamma(\epsilon)\Gamma(-\epsilon)\Gamma^2(2\epsilon)\Gamma(1-2\epsilon) +\pmb B^{2\epsilon}\frac{\Gamma^2(-\epsilon)}{2\epsilon}\(\mathbf{L}_{\delta}-\mathbb{S}^1_{\epsilon}-\mathbb{S}^1_{2\epsilon}+\mathbb{S}^1_{2\epsilon-1}\)
\Bigg].
\end{eqnarray}
where we have introduced notations
\begin{eqnarray*}
F_\epsilon&=&2^{2-\epsilon}\frac{\Gamma^2(-\epsilon)\Gamma^2(1+\epsilon)}{\epsilon^2}\,_2F_1(1,1,1+\epsilon;-1),
\\
Q(\epsilon)&=&\sum_{k=1}^\infty \frac{(-1)^k}{k!}\Gamma(k-\epsilon)\(\Gamma(k)\Gamma(-k-\epsilon)\psi\(\frac{1+k}{2}\)+\Gamma(k-2\epsilon)\Gamma(\epsilon-k)\psi\(\frac{1+k-\epsilon}{2}\)\)
\\&=&-\frac{23}{16}\zeta_4-\zeta_2 \ln^2 2-2\gamma_E\zeta_3+\frac{\ln^4 2}{6}+4 \Li_4\(\frac{1}{2}\)-\frac{\ln 2}{2}\zeta_3+\mathcal{O}(\epsilon)
\end{eqnarray*}
The integral $I''_A$ has been evaluated in \cite{Echevarria:2015byo}. Integrals $I_{C1,C2}$ related to those calculated in \cite{Echevarria:2015byo} as $I_{C1}(\vec 0)=I_{C1}$, $I_{C2}(\vec 0)=I_{C2}$, $I_{C1}(\vec b)=-I''_{C4}$ and $I_{C2}(\vec b)=I''_{C3}$.

\subsection{Three point integrals}

Three point integrals appear in the diagrams that connect three Wilson lines and contribute into $W_{ijk}$. Generally, these integrals cannot be presented in the closed form.

The integral that appears in diagrams of group ${\langle 211\rangle}$ is
\begin{eqnarray}\label{app:Int_J}
&&J(\vec b_1,\vec b_2)=
\int \frac{d^dkd^dl}{(2\pi)^{2d}}
\frac{-e^{i\vec k\vec b_1}e^{i\vec l\vec b_2}}{k^2l^2(k^++i\delta^+)(l^++i\delta^+)(k^-+l^--2i\delta^-)(l^--i\delta^-)}
\\\nn &&= \frac{2}{(4\pi)^d}
\Bigg[\pmb \delta^{-2\epsilon}\Gamma^4(\epsilon)\Gamma^2(1-\epsilon)+
(\pmb B_1^\epsilon-\pmb B_2^\epsilon)\pmb \delta^{-\epsilon}F_\epsilon
\\&& \nn\quad
+(\pmb B_1 \pmb B_2)^{\epsilon}\Gamma^2(-\epsilon)\((\mathbf{L}_{\delta1}-\mathbb{S}^1_{\epsilon})^2-\mathbb{S}^2_{\epsilon}+3\zeta_2\)
-2\pmb B_1^\epsilon \pmb \delta^{-\epsilon}\Gamma(-\epsilon)\Gamma^2(\epsilon)\Gamma(1-\epsilon)\(\mathbf{L}_{\delta 1}-\mathbb{S}^1_{\epsilon}\)
\\
&&\nn\quad
+2\pmb B_1^{2\epsilon}\frac{\Gamma^2(-\epsilon)\Gamma(-2\epsilon)\Gamma^2(1+\epsilon)}{\epsilon}
+2(\pmb B_1 \pmb B_2)^\epsilon R\(\frac{\vec b_1^2}{\vec b_2^2}\)
\Bigg],
\end{eqnarray}
where $\mathbb{L}_{\delta 1}=\ln(\pmb \delta \pmb B_1e^{2\gamma_E})$, and
\begin{eqnarray}\nn
R(x)&=&\int_{-i\infty}^{i\infty} \frac{ds}{2\pi i} \frac{\Gamma^2(-s)\Gamma^2(1+s)}{s}\Gamma(-s-\epsilon)\Gamma(s-\epsilon)x^s,
\end{eqnarray}
where the integration contour passes between series of poles. 

The integrals $J$ with exchanged variables are related to each other
\begin{eqnarray}
J(\vec b_1,\vec b_2)+J(\vec b_2,\vec b_1)=K^{(0)}(\vec b_1)K^{(0)}(\vec b_2),
\end{eqnarray}
which can be checked explicitly using expression (\ref{app:Int_J}). If one of the variables zero, the expression can be simplified
\begin{eqnarray}
J(0,\vec b)&=&\frac{2}{(4\pi)^d}\Bigg[\pmb \delta^{-2\epsilon}\Gamma^4(\epsilon)\Gamma^2(1-\epsilon)
-\pmb B^\epsilon \pmb \delta^{-\epsilon}F_\epsilon
-2\pmb B^{2\epsilon}\Gamma^2(-\epsilon)\Gamma(-2\epsilon)\Gamma(\epsilon)\Gamma(1+\epsilon)\Bigg].
\end{eqnarray}
In this limit the integral has been evaluated in \cite{Echevarria:2015byo} with the same result, which can be seen by relation $J(\vec 0,\vec b)=-I_A'$.

The integral that appears in diagram ${\langle 111\rangle}$ is
\begin{eqnarray}\label{app:Int_R2}
R(\vec b_1,\vec b_2)&=&\int \frac{d^dkd^dl}{(2\pi)^{2d}}\frac{-e^{i\vec k\vec b_1}e^{i\vec l\vec b_2}}{(k^++i\delta^+)(l^-+i\delta^-)k^2l^2(k+l)^2}=R^{sing}(\vec b_1,\vec b_2)+R^{reg}(\vec b_1,\vec b_2).
\end{eqnarray}
The $R^{sing}$ is a subintegral that contains the soft singularities and double-rapidity singularities and can be evaluated explicitly
\begin{eqnarray}
R^{sing}(\vec b_1,\vec b_2)&=&\frac{2}{(4\pi)^d}\Bigg[\pmb B_1^\epsilon \pmb B_2^\epsilon \frac{\Gamma^2(-\epsilon)}{2}\((\mathbf{L}_{\delta 2}-\mathbb{S}^1_{\epsilon}+\mathbb{S}^1_{-\epsilon-1})^2-\mathbb{S}^2_{\epsilon}-\mathbb{S}^2_{-\epsilon-1}+6\zeta_2\)
\\&&\nn-\pmb B_2^{2\epsilon}\Gamma(-\epsilon)\Gamma(-2\epsilon)\Gamma(\epsilon)\(\mathbf{L}_{\delta 2}-\mathbb{S}^1_{\epsilon}-\mathbb{S}^1_{2\epsilon}+\mathbb{S}^1_{-\epsilon-1}\)
\\&&\nn-\pmb \delta^{-\epsilon}\(\pmb B_1^\epsilon+\pmb B_2^\epsilon\)\Gamma^2(-\epsilon)\Gamma^2(\epsilon)+\pmb \delta^{-2\epsilon}\Gamma(-\epsilon)\Gamma(\epsilon)\Gamma(2\epsilon)\Gamma(-2\epsilon)\Gamma(1+2\epsilon)
\Bigg].
\end{eqnarray}
The regular part contains only single rapidity divergence, and has the following integral representation
\begin{eqnarray}
&&R^{reg}(\vec b_1,\vec b_2)=\frac{2}{(4\pi)^d}\int_0^1 \frac{dy}{y}\Bigg[-\pmb B_1^\epsilon \Gamma^2(-\epsilon)\(\pmb B_{21,y}^\epsilon-(\bar y \pmb B_2)^\epsilon\)\(\mathbf{L}_{\delta 2}-\mathbb{S}^1_{\epsilon}-\ln(y/\bar y)\)
\\&&\nn -(y\bar y)^{-\epsilon}\frac{\Gamma(-2\epsilon)}{\epsilon}\(\pmb B_{21,y}^{2\epsilon}\,_2F_1\(-\epsilon,-2\epsilon,1-\epsilon;\frac{-y\bar y \pmb B_1}{\pmb B_{21,y}}\)-(\bar y \pmb B_2)^{2\epsilon}\)\(\mathbf{L}_{\delta 2}-\mathbb{S}^1_{2\epsilon}-\ln(y/\bar y)\)
\\&&\nn-\pmb B_1^\epsilon \pmb B_{21,y}^\epsilon \Gamma^2(-\epsilon)\ln \(\frac{\pmb B_{21,y}}{\bar y \pmb B_2}\)
-(y\bar y)^{-\epsilon}\frac{\Gamma(-2\epsilon)}{\epsilon}\pmb B_{21,y}^{2\epsilon}
\,_2F_1\(-\epsilon,-2\epsilon,1-\epsilon;\frac{-y\bar y \pmb B_1}{\pmb B_{21,y}}\) \ln \(\frac{\pmb B_{21,y}}{\bar y \pmb B_2}\)
\\\nn&&+(y\bar y)^{-\epsilon}\frac{\Gamma(-2\epsilon)}{\epsilon}\pmb B_{21,y}^{2\epsilon}
\,_2F_1^{(0,1,0)}\(-\epsilon,-2\epsilon,1-\epsilon;\frac{-y\bar y \pmb B_1}{\pmb B_{21,y}}\)\Bigg],
\end{eqnarray}
where $\pmb B_{21,y}=(\vec b_2-y \vec b_1)^2/4$, and $\,_2F_1^{(0,1,0)}$ is the derivative of hypergeometric function over the second index (it can be expressed as $\,_3F_2$-function). The integral over $y$ is regular at $\epsilon\to 0$, and can be integrated order-by-order of $\epsilon$-expansion. One can check that $R(\vec b_1,\vec b_2)=R(\vec b_2,\vec b_1)$.

Within our calculation we do not need the complete expression for $R$ but only its limiting cases,
\begin{eqnarray}
R(\vec b,\vec b)&=&\frac{-2}{(4\pi)^d}\Bigg[
-2\pmb B^{\epsilon}\pmb \delta^{-\epsilon}\Gamma^2(\epsilon)\Gamma^2(-\epsilon)-\pmb \delta^{-2\epsilon}\Gamma^2(2\epsilon)\Gamma(\epsilon)\Gamma(-\epsilon)\Gamma(1-2\epsilon)
\\&&\nn+\pmb B^{2\epsilon}\Gamma^2(-\epsilon)\(\frac{1}{2}\(\mathbf{L}_\delta +\mathbb{S}^1_{2\epsilon}+\mathbb{S}^1_{-\epsilon-1}-2\mathbb{S}^1_{\epsilon}\)^2+\frac{3}{2}\mathbb{S}^2_{2\epsilon}-\mathbb{S}^2_{\epsilon}-\frac{1}{2}\mathbb{S}^2_{-\epsilon-1}+2\zeta_2\)
\Bigg],
\\
R(\vec b,\vec 0)&=&\frac{2}{(4\pi)^d}\Bigg[
\pmb B^\epsilon \pmb \delta^{-\epsilon}\Gamma^2(\epsilon)\Gamma^2(-\epsilon)+\pmb\delta^{-2\epsilon}\Gamma(\epsilon)\Gamma(-\epsilon)\Gamma^2(2\epsilon)\Gamma(1-2\epsilon)
\\\nn&&+\pmb B^{2\epsilon}\frac{\Gamma^2(-\epsilon)}{2\epsilon}(\mathbf{L}_\delta-\mathbb{S}^1_{\epsilon}-\mathbb{S}^1_{2\epsilon}+\mathbb{S}^1_{2\epsilon-1})\Bigg].
\end{eqnarray}
These cases can be related to the integrals evaluated in \cite{Echevarria:2015byo} as $R(\vec b,\vec b)=I'_{C3}+I''_{C1}$, $R(\vec b,\vec 0)=-I''_{C2}$ and $R(\vec 0,\vec 0)=-I_{C2}$.

\end{document}